\documentclass[authoryear,3p,11pt]{elsarticle}

\makeatletter
\def\ps@pprintTitle{%
	\let\@oddhead\@empty
	\let\@evenhead\@empty
	\let\@evenfoot\@empty
	\let\@evenfoot\@empty	
	}
\makeatother

\usepackage{epstopdf}
\usepackage{amsthm,amsmath}
\usepackage{amssymb}
\usepackage{algorithm}
\usepackage{algpseudocode}
\usepackage{algorithmicx}
\usepackage{graphicx}
\usepackage{mathrsfs}
\usepackage{booktabs}
\usepackage[colorlinks,citecolor=blue,urlcolor=blue]{hyperref}
\usepackage{pstool}
\usepackage[caption=false]{subfig}
\usepackage{tikz}
\usepackage[colorlinks,citecolor=blue,urlcolor=blue]{hyperref} 

\graphicspath{{figs_png/}} 

\def\v{{\varepsilon}}

\def\argmax{\mathop{\rm argmax}}
\def\argmin{\mathop{\rm argmin}}

\def\trace{\mathop{\rm tr}}
\def\det{\mathop{\rm det}}

\def\logm{\mathop{\rm logm}}

\def \e {{e}}

\def \p{{p}}

\def \v {{v}}
\def \x{{x}}
\def \y{{y}}

\def \z{{z}}

\newcommand{\real}{\ensuremath{\mathbb{R}}}
\newcommand{\s}{\ensuremath{\mathbb{S}}}

\begin{document}

\begin{frontmatter}
\author[nswc]{M. Rosenthal}
\ead{michaelr@stat.fsu.edu}
\author[statfsu]{W. Wu}
\ead{wwu@stat.fsu.edu}
\author[mathfsu]{E. Klassen,}
\ead{klassen@math.fsu.edu}
\author[statfsu]{Anuj Srivastava}
\ead{anuj@stat.fsu.edu}
\address[nswc]{Naval Surface Warfare Center, Panama City Division - X23, 110 Vernon Avenue, Panama City, FL 32407-7001}
\address[statfsu]{Department of Statistics, Florida State University, Tallahassee, FL 32306}
\address[mathfsu]{Department of Mathematics, Florida State University, Tallahassee, FL 32306}
\title{Nonparametric Spherical Regression Using Diffeomorphic Mappings}

\begin{abstract}
Spherical regression explores relationships between variables on spherical domains. 
We
develop a nonparametric model that uses a diffeomorphic map from a sphere to itself. 
The restriction of this mapping to diffeomorphisms is natural
in  several settings.
The model is estimated in a penalized maximum-likelihood framework using 
gradient-based optimization.  Towards that 
goal, we specify a first-order roughness penalty using the Jacobian of diffeomorphisms. 
We compare the prediction performance of the proposed model with state-of-the-art methods using simulated and 
real data involving cloud deformations, wind directions, and vector-cardiograms. This model is found to outperform
others in capturing relationships between spherical variables. 
\end{abstract}

\end{frontmatter}
\section{Introduction}
Spherical data arises naturally in a variety of settings. 
For instance, a random vector with unit norm constraint is naturally studied as a point on a unit sphere.
The statistical analysis of such random variables
was pioneered by Mardia and colleagues
\citeyearpar{mardia:1972,jupp-mardia:2000}, in the context of directional data. 
Common application areas where such data originates include geology,  gaming, meteorology, computer vision, 
and bioinformatics. 
Examples from geographical domains include plate tectonics \citep{mcKenzie:1957,chang:1986}, animal migrations, and tracking of weather formations. As mobile devices  become increasingly advanced and prevalent, an abundance of new spherical data is being collected in the form of geographical coordinates.  Another source of spherical data studies directions, 
e.g., 
vector-cardiograms studied in \cite{downs:2003}. Directional data also characterizes the orientations of objects or 
limbs, 
which are particularly relevant to biometric applications including the study of human kinematics \citep{Rancourt:2000} and gait data in the context gait identification \citep{Boyd:2005}.

Spherical regression is an analysis of paired data on a unit hyper-spherical domain $\mathbb{S}^{d-1}=\{\z\in \mathbb{R}^d: \|\z\|=1 \}$, where $\| \cdot \|$ indicates the 
Euclidean 
norm.  Given $n$ paired observations $(\x_i,\y_i) \in \mathbb{S}^{d-1}\times \mathbb{S}^{d-1}$ for $i=1,\ldots, n$, one wishes to describe the relationship between predictor $\x$ and response $\y$ in order to make predictions and inferences. To define a spherical regression model, one must decide on the functional form of $\mu(\x)$ the mean of $\y$ for a given $\x$. This function $\mu:\mathbb{S}^{d-1} \rightarrow \mathbb{S}^{d-1}$ characterizes the expected relationship between the $\x$ and $\y$ variables, and can take a parametric, semi-parametric, or a nonparametric form. The estimation of $\mu$ additionally depends on the chosen spherical error distribution for $\y$ given $\x$.  

In past work,
the function $\mu$ has predominantly taken a parametric form. 
Several parametric models have been proposed for the unit circle $\mathbb{S}^1$,  including the one in \cite{rivest:1997} and \cite{downs-mardia:2002}, 
but the choice gets limited for higher dimensions.  There,  the rigid rotation model 
has been the most common choice \citep{chang:1986,chang:1989,rivest:1989,kim:1991,prentice-mardia:1995}. 
 \cite{downs:2003} 
used complex M\"obius transformations on $\s^2$. More recently,   \cite{PLT:2012} expanded
the parametric class by including projective linear transformation. 
Additionally, there has also been some progress in nonparametric formulations. 
For instance,  \cite{Taylor:2013} used a model that  modifies standard kernel-smoothing methods to derive a general 
nonparametric spherical regression model.

Parametric models, especially 
 with a small number of parameters, are often too restrictive to adequately capture broad correspondences observed in real data. On the other hand, the high dimensionality of certain nonparametric models can lead to 
overfitting and, thus, hinder model performance. Our proposed 
framework
includes
 flexible classes of deformations using diffeomoprhic 
nonparametric representations 
which avoid over-fitting by using penalty functions.
Smooth bijective relationships 
%
occur naturally in many systems. An example is fluid 
motion, as 
the mass can be compressed or expanded but is not generally allowed to occupy the same space simultaneously. For the same reason, 
the relationships between atmospheric variables over short intervals  can be modeled using diffeomorphisms.  Since diffeomorphisms are invertible mappings, they are also useful in situations where the roles of $\x$ and $\y$ can be reversed. In this paper, we take a nonparametric approach using a diffeomorphism from $\s^{2}$  to itself. 
The technique we develop are restricted to $\s^2$  due to the optimization procedure we have used, although
the underlying penalized-likelihood framework itself is valid for arbitrary dimensions.



\section{Previous Methods and Their Limitations}\label{sec:lit_review}
\subsection{Parametric Methods:}
A majority of past 
work
in spherical regression have imposed parametric forms for $\mu$. \cite{chang:1986,chang:1989} created this field in the 80's using $\mu(\x)=A\x$ where $A $ is a rotation 
 in $\mathrm{SO}(d) =\{ A\in \mathbb{S}^{d\times d}: \det(A)=1, A^TA=AA^T=I_d\}$.
This approach was 
%
followed by 
%
\cite{rivest:1989} and \cite{kim:1991}. The solution which minimizes the sum of squared errors is given by \cite{mcKenzie:1957} and \cite{stephens:1979} as the Procrustes rotation.  The rigid rotation model can characterize changes in location and orientation, but 
%
is unable to handle any difference in spread between the predictor and response variable. It is akin to using the fixed addition model $\y= (\x + a)+ \e$ in the classical 
Euclidean 
case.

\cite{downs:2003} proposed an extension that uses M\"obius transformations, a larger parametric family than rotations.  This model can handle some additional differences in spread between the response and predictor variable, but this framework has not been extended to dimensions higher than two. 
A recent model that is applicable to arbitrary $\mathbb{S}^{d-1}$ uses the group of projective linear transformations.  
A projective linear transformation is a map $\x \rightarrow A/|A\x|$ and is parametrized by the transformation matrix 
$A\in \mathrm{SL}(d)$, the set of 
all matrices with determinant one.  In \cite{PLT:2012}, the authors derive an intrinsic
Newton Raphson
algorithm for maximum-likelihood estimation and present an asymptotic analysis of the estimator under the von 
Mises Fisher
error distribution. 

These parametric models are likely to be useful for a variety of inference and prediction applications. However, more general situations require more flexible models in order to adequately characterize relationships between spherical variables. 
While one 
possibility is to explore
even larger classes of parametric families, a more natural approach is nonparametric regression.

\subsection{Nonparametric Methods:}
Nonparametric 
methods allow flexible expressions of $\mu$ at the cost of model parsimony.
In recent work, \cite{Taylor:2013} developed a nonparametric regression model that locally fits  weighted polynomials at each point on the sphere. 
They demonstrate improvements in predictive performance over simpler kernel smoothing and 
%
rigid rotations. This model is applicable across arbitrary hyper-spheres, dimensions, and geometries. For example, 
it
can be used to define relationships between spherical and linear domains. As with most kernel smoothing approaches, 
it
does not ensure invertibility of regression maps. This limitation becomes evident in applications such as meteorological studies, where the short term changes are often diffeomorphic, and 
where data is often scarce over large areas of the sphere. 
This
is a common occurrence with data generated from satellites. In these cases, kernel smoothing methods can overfit data 
in ways that the bandwidth or smoothing parameter cannot adequately address. A restriction to the group of diffeomorphisms is a 
natural
alternative in such settings. 

Our proposed approach is similiar to \cite{Glaunes:2004} which utilizes diffeomorphic maps on the sphere in the context of landmark matching. 
They define an objective function for landmark-guided point registration, but this function is not motivated from a regression model perspective. Thus, their approach is not directly applicable to prediction and model estimation. 
The spherical regression setting is additionally concerned with prediction accuracy, model parsimony, interpretability, and statistical inference. Our approach will 
use
a different objective function, roughness penalty, and optimization strategy that
are
more suitable for spherical regression.

\section{Proposed Regression Framework}\label{sec:our_framework} 

\subsection{Penalized Maximum-Likelihood Estimation}

The proposed method characterizes relationships between spherical data on $\mathbb{S}^{d-1}$ using diffeomorphisms. The set of such diffeomorphisms $\Gamma $ consists of mappings $\gamma:\mathbb{S}^{d-1} \rightarrow\mathbb{S}^{d-1}$ that are smooth and invertible with smooth 
inverses, and
forms a group under composition. The identity element of 
%
$\Gamma$ is the mapping $\gamma_{id}(\x)=\x$. The group $\Gamma$ forms a very flexible class of deformations which includes the subgroup of rigid rotations and the projective linear group.  Additionally, the complex Mobius maps, considered as maps on $\mathbb{S}^2$, are diffeomorphisms of the Riemann sphere. For more details on diffeomorphisms see \cite{kac-book}.

We will need elements of the Riemannian geometry of $\mathbb{S}^{d-1}$  and $\Gamma$ to derive 
%
estimation algorithms.
The tangent space at 
%
$\p \in \mathbb{S}^{d-1}$ is denoted by $T_{\p}(\s^{d-1})= \{\x\in\mathbb{R}^d: \langle \x,\p\rangle=0 \}$ and is a 
vector space of dimension $d-1$.
Similarly, the tangent space of $\Gamma$ at $\gamma_{id}$ is denoted by
 $T_{\gamma_{id}}(\Gamma)$ and is infinite dimensional. This tangent space  $T_{\gamma_{id}}(\Gamma)$ is the set of smooth tangent vector fields on $\s^{d-1}$.  That is, for any $V\in T_{\gamma_{id}}(\Gamma)$ and  $\p\in \mathbb{S}^{d-1}$, $V(\p)  \in T_\p(\mathbb{S}^{d-1})$ is  a tangent vector, and it varies smoothly with $\p$.

We assume that the error distribution of $\y$ given $\x$ follows a 
{ von 
Mises Fisher distribution} with mean direction $\gamma(\x)$ ($\gamma \in \Gamma$) and concentration parameter $\kappa>0$.
That is,
\begin{equation}
p_\x({\y}) = C_d(\kappa)  \exp\{\kappa  {\y}^T \gamma (\x)\}\ ,
C_d(\kappa) =  { \kappa^{d/2 -1} \over (2 \pi)^{d/2} \mathcal{I}_{d/2-1}(\kappa)} ,\\
\label{eqn:vmf-density}
\end{equation}
with 
${\y}, \x, \gamma(\x) \in \mathbb{S}^{d-1}$.
Here, $C_d$ is the normalizing constant, and
$\mathcal{I}_\nu$  is the modified Bessel function of the first kind and order $\nu$.
The von 
Mises Fisher   
density is isotropic about its mean direction $\gamma(\x)$. The parameter $\kappa$ measures the degree of 
concentration:
$\kappa = 0$ implies a uniform density on $\s^{d-1}$ while $\kappa = \infty$ implies a Dirac delta at $\gamma(\x)$.  For further details 
%
see \cite{jupp-mardia:2000}. 
The normalizing constant $C_d(\kappa)$ does not depend on the mean direction, so maximum likelihood estimation of $\gamma$  separates from that of $\kappa$

Since the group of diffeomorphisms is an infinite-dimensional function space, 
the maximum likelihood estimate may over-fit the data.  
To overcome this problem, we 
seek a penalized maximum log-likelihood solution solution that takes the form
$$\hat{\gamma}=
\argmax_{\gamma\in \Gamma} \left( n^{-1}\sum_{i=1}^n \y_i^T \gamma(\x_i) - \lambda R(\gamma)\right)\ .$$
The term $f_n(\gamma)=n^{-1}\sum_{i=1}^n \y_i^T \gamma(\x_i)$ is the log-likelihood 
%
from equation (\ref{eqn:vmf-density}), and  the function $R:\Gamma\rightarrow \mathbb{R}_+$ measures the roughness of $\gamma$.  
We introduce
a scale $n^{-1}$ in the log-likelihood term in order to compare it across sample sizes. 
The scalar $\lambda>0$ denotes a tuning parameter which controls the amount of penalty for the roughness term. In the next section, we will construct a specific $R(\gamma)$.

\iftrue
\subsection{Two Roughness Penalties Based on Distance from Isometry}\label{sec:roughness}
It is common in nonparametric statistics involving 
Euclidean 
variables to use first or second derivatives of functions to define their roughness.
The first order penalty is zero for translations and penalizes large slopes. The second order penalty is zero for linear maps and 
penalizes curvatures. The interpretation of transformations in spherical domains is
%
different.  
Translations in the
Euclidean 
 domain become rigid rotations and reflections in spherical domains, but the analogs of 
higher-order transformations are not clear. 
%
We require that if
the diffeomorphism is either a rotation or a reflection, i.e. $\gamma(\p) = O\p$, 
$O \in \mathcal{O}(d)=\{O\in \real^{d\times d}: O^TO=OO^T=I_d\}$, with $I_{d}$ denoting the $d\times d$ identity matrix ,
then its roughness measure should be zero. For this purpose, it is sufficient to impose a penalty that uses
the first derivative, or the Jacobian, of $\gamma$. 
Not only do we not have a proper interpretation for the second derivative of 
$\gamma$, 
but it also becomes computationally complex. Consequently, we will derive a first-order roughness penalty on $\gamma$.

Let $J_\p(\gamma)$  denote the Jacobian of $\gamma\in \Gamma$ evaluated at a point $\p \in\s^{d-1}$. By definition,
$J_\p(\gamma)$ 
is a linear mapping from the tangent space $T_\p(\s^{d-1})$ to $T_{\gamma(\p)}(\s^{d-1})$, each one of these tangent spaces being $(d-1)$-dimensional.
Therefore, we can express the Jacobian as a $(d-1)\times (d-1)$ matrix with respect to the chosen bases for these tangent spaces, as elaborated next.
Any linear map between the two tangent spaces can first be expressed in $\real^d$ coordinates as
$u \mapsto B u$ 
for $B \in \mathbb{R}^{d\times d}$ and each 
${u} \in T_\p(\s^{d-1})$ 
viewed as an element of $\real^d$.
 Let 
$E,F \in \real^{d \times (d-1)}$ 
denote
orthonormal bases for $T_\p(\s^{d-1})$ and $T_{\gamma(\p)}(\s^{d-1})$, respectively. Then, under the chosen
bases, the Jacobian matrix becomes
$J_\p = F^T B E \in \real^{(d-1) \times (d-1)}$. 
In case $\gamma(\p) = O\p$ for an $O \in \mathcal{O}(d)$, then $OE \in \real^{d \times (d-1)}$ also forms
an orthonormal basis of $T_{\gamma(\p)}(\s^{d-1})$.
Since $F$ and $OE$ are two orthonormal bases of the same space, there exists an orthogonal matrix $A \in \mathcal{O}(d-1)$ such that
$F = OEA$. Thus, the expression for the Jacobian in this case reduces to
%
$J_p = F^T OE = A^T E^T O^T O E = A^T \in \mathcal{O}(d-1)$.

Based on this discussion, we can define roughness
at point $\p$ is $\| J_{\p}(\gamma)^TJ_{\p}(\gamma)-I_{d-1}\|^2$, where $\|\cdot \|$ denotes the Frobenius matrix norm.
The roughness at $\p$ is zero if and only if $J_{\p}(\gamma)$ is an orthogonal matrix, i.e., a rotation or a reflection.
The resulting first-order roughness measure for the full map $\gamma$ is then:
$Q(\gamma)= \int_{\mathbb{S}^{d-1}}\| J_{\p}(\gamma)^TJ_{\p}(\gamma)-I_{d-1}\|^2d\p$.
Another interesting property of this definition is that the natural action of the subgroup $\mathcal{O}(d)$ on $\Gamma$
leaves the roughness measure unchanged. That is, for any $\gamma \in \Gamma$ and $O \in \mathcal{O}(d)$, if we
define a new $\tilde{\gamma}(\p) = O \gamma(\p)$, then $Q(\tilde{\gamma}) = Q(\gamma)$. 
\fi

Another measure of the distance from isometry can be computed as $\|\logm(J_{\p}^TJ_{\p})\|^2$ where $\logm$ denotes the matrix log and $\|\cdot\|$ denotes the matrix Frobenius norm. Unlike the previous distance from isometry measure, this will diverge to infinity as the Jacobian becomes singular. The resulting roughness measure is defined as $R(\gamma)= \int_{\mathbb{S}^{d-1}}\| \logm\{J_{\p}(\gamma)^TJ_{\p}(\gamma)\}\|^2d\p$. 
This has the added property that the roughness measure will be infinity if the Jacobians are singular over a set of positive measure. 
As a penalty term, this will ensure that the deformation's Jacobean has nonzero determinant everywhere except perhaps on a set of measure zero. 

Because of the group structure of $\Gamma$, any finite sequence of composed diffeomorphisms will result in a diffeomorphism. In the limit, an infinite sequence of compositions may converge to a non-diffeomorphic map. For example, a set of points with positive measure may converge to a set of measure zero in the limit. 
On the other hand, the proposed roughness measure of such a deformation is infinity, so the roughness measure should push the deformation away from such solutions.
In this paper, we will use $R$ as our chosen roughness measure. Details for computing these roughness measures $Q$ and $R$
are presented in the Appendix.
\section{Optimization Algorithm} \label{sec:estimation}
\subsection{Algorithm Overview}
Returning to the problem of model estimation, we treat it as an optimization
of the objective function
$E(\gamma) = n^{-1} \sum_{i=1}^n \y_i^T \gamma(\x_i) - \lambda R(\gamma)$ by 
gradient ascent.
Since we are optimizing over a nonlinear group, the Riemannian geometry of $\Gamma$
plays a vital role. The gradient of a function at a point on the Riemannian manifold is by definition an element of the tangent space at that point. Since tangent spaces are linear, one can represent their elements as coefficients with respect to 
corresponding orthonormal bases. Thus, the gradient of 
$E$ at any $\gamma$ 
can be expressed as a linear combination of tangent basis elements at that $\gamma$. 
This simplifies the computation of gradient to solving for the corresponding coefficients, but it still
requires an orthonormal basis of all tangent spaces of $\Gamma$. To avoid this, we will take an iterative approach and solve of an optimal incremental 
diffeomorphism at every iteration as follows.

We 
%
define an incremental cost function $H(\tilde{\gamma}) = E(\tilde{\gamma} \circ {\gamma})$ for 
the current estimate $\gamma \in \Gamma$. 
The incremental diffeomorphism $\tilde{\gamma}$ will be small,
 i.e.,
 close to $\gamma_{id}$, and can be related to an element of 
 $T_{\gamma_{id}}(\Gamma)$. Thus, we need to specify the tangent space at $\gamma_{id}$ only and,  as mentioned 
earlier, this is a set of all smooth tangent vector fields on $\s^d$.
The gradient of $H$ at $\gamma_{id}$, $\nabla_{\gamma_{id}} H$, is an element of $T_{\gamma_{id}}(\Gamma)$. 
Given an orthonormal basis 
$\{B_1,B_2,\ldots\}$ 
of $T_{\gamma_{id}}(\Gamma)$, 
we can write  
$\nabla_{\gamma_{id}} H= \sum_{j=1}^{\infty} d_j B_j$,
where each $d_j\in \mathbb{R}$. 
Each $d_j$ coefficient is the directional derivative of $H$ in the direction of $ B_j$ at $\gamma_{id}$. 
In Section \ref{Orthogonal Basis of Spherical Deformations} we define such a basis for the tangent space $T_{\gamma_{id}}(\Gamma)$. Then in Section \ref{Gradient Ascent Algorithm} we show how to compute the gradient and 
present
the gradient algorithm.

Additionally, we will need the following tools on the sphere $\s^d$. 
The exponential map at $\p\in \s^{d-1}$ is a mapping $\exp_\p:T_\p(\s^{d-1})\rightarrow \s^{d-1}$
according  to
$\exp_\p(\v)= \cos(\|\v\|)\p +\sin(\|\v\|)\v/\|\v\| \in \s^{d-1}$.
The inverse exponential map at point $\p \in \mathbb{S}^{d-1}$ maps each non-antipodal point  $\z\in \mathbb{S}^{d-1}$ to a tangent vector $\v \in T_\p(\mathbb{S})$ according to
$\exp^{-1}_{\p}(\z)=(\z-\cos(\theta)\p) \theta (\sin(\theta))^{-1}$, where 
$\theta=\cos^{-1}\left(\p^T\z\right)$. 
Two points $\p,\z\in \s^{d-1}$ are non-antipodal if $\p^T\z\neq -1$.
For $V\in T_{\gamma_{id}}(\Gamma)$ we define a map expressed as $\Psi_V(\p)= \exp_{\p}(V(\p))$ for each $\p \in \mathbb{S}^{d-1}$.
The map  $\Psi_V$ is a diffeomorphism on $\s^{d-1}$ when $V$ is in a small neighborhood around the 
 zero vector. The zero vector is denoted by the vector field $V_0$ with $\|V_0(\p)\|=0$ for each $\p \in \s^{d-1}$.

\subsection{Orthogonal Basis for Incremental Diffeomorphisms}\label{Orthogonal Basis of Spherical Deformations}
In this paper, we focus on $d=3$ and develop the optimization algorithm for that case.
In \cite{kurtek-et-all:11}
%
construct an orthonormal basis of smooth tangent vector fields on a sphere for $T_{\gamma_{id}}(\Gamma)$ 
by applying the gradient to the real and imaginary parts of the complex spherical harmonic function $Y_l^m$ of degree $l$ and order $m=0,\ldots, l$. 
From the first $l$ harmonics one gets $(l+1)^2$ distinct functions by taking the real and imaginary parts
as separate functions.  We will denote these 
real-valued 
functions on $\s^2$ by $\varphi_1,\ldots, \varphi_{(l+1)^2}$ and 
note that they are parametrized using polar coordinates. For each 
$i \in 1,\ldots ,(l+1)^2$ 
we evaluate the gradient at each point $(\theta,\phi) \in \s^2$ 
resulting in the vector field 
$$\nabla_{(\theta,\phi)} \varphi_i=\left(\frac{\partial\varphi_i}{\partial\theta}, \frac{1}{\sin(\theta)}\frac{\partial\varphi_i}{\partial\phi}\right)\ .$$ 
For each resulting non-trivial tangent vector field $\nabla \varphi_i$ for $i \in \{1,2,\ldots\}$,  let $\tilde B_i(\theta,\phi) =\nabla_{(\theta,\phi)} \varphi_i / \|\nabla_{(\theta,\phi)} \varphi_i\|$. Let $*\tilde B_i(\theta,\phi)$ denote the tangent vector at point $(\theta,\phi)$ obtained by rotating $\tilde B_i(\theta,\phi)$  counterclockwise by $\pi/2$ in its
tangent space. According to 
{Proposition 3} of \cite{kurtek-et-all:11}, the union of these non-trivial tangent vector fields  $\{\tilde B_i\}_{i=1,2,\ldots}$ and the rotated tangent vector fields  $\{*\tilde B_i\}_{i=1,2,\ldots}$ provides an orthonormal basis for $T_{\gamma_{id}}(\Gamma)$. 

We denote this union by $\{ B_1,B_2,\ldots\}$. 
For any $B \in T_{\gamma_{id}}(\Gamma)$ and any point $\p \in \s^2$, the tangent vector  $B(\p) \in T_\p(\s^2)$ is represented as an element of $\mathbb{R}^3$.
If we restrict to basis elements obtained from spherical harmonics of order $\leq l$, we obtain $L=2(l+1)^2-2$ distinct non-trivial basis elements. As  $l$ increases, 
so does 
%
$L$, and one is able to capture more complex deformations.

\subsection{Gradient Ascent Algorithm}
\label{Gradient Ascent Algorithm}
As mentioned earlier, each iteration is based on optimization of 
the functional  $H(\tilde{\gamma}) = E(\tilde{\gamma} \circ \gamma)$ where $\gamma\in \Gamma$ is the current deformation, 
and we optimize over the increment $\tilde{\gamma}$ in the neighborhood of $\gamma_{id}$. 
By definition, $\nabla_{\gamma_{id}}H \in T_{\gamma_{id}}(\Gamma)$, which implies that the gradient of $H$ at the identity can be expressed as a linear combination 
$\nabla_{\gamma_{id}} H= \sum_{j=1}^{\infty} d_j B_j$.  Each coefficient $d_j$ is the directional derivative of $H$ in the direction of $B_j\in T_{\gamma_{id}}(\Gamma)$ given by
\begin{align*}
d_j=
\lim_{\epsilon \downarrow 0} 
\frac{
H (\Psi_{\epsilon B_j}) - H(\gamma_{id})
}{\epsilon}=
\lim_{\epsilon \downarrow 0} 
\frac{
E (\Psi_{\epsilon B_j}\circ \gamma) - E(\gamma)
}{\epsilon} \ .
\end{align*}

Since the gradient operation is linear, we can separate  
the
likelihood and roughness 
terms in $H$. We can write an analytical expression for the directional derivatives of the log-likelihood term.
Let $f_{(i)}(\gamma) = \y_i^T\gamma(\x_i)$ and ${\z}_i=  \gamma(\x_i)$, then the coefficient $a_{ij}\in \mathbb{R}$ is computed as follows: 

\begin{align*}
a_{ij}
&=\lim_{\epsilon \downarrow 0}\frac{
f_{(i)}(\Psi_{\epsilon  B_j} \circ  \gamma) -f_{(i)}(\gamma)}{\epsilon}
=\lim_{\epsilon \downarrow 0}\frac{\y_i^T\exp_{\z_i}(\epsilon   B_j( \z_i)) -\y_i^T\z_i}{\epsilon}
\notag\\
 &=
 \y_i^T \left.\left(
-\sin(\epsilon\|   B_j(\z_i)\|)\z_i\|   B_j(\z_i)\| +\cos(\epsilon\|  B_j(\z_i)\|)   B_j(\z_i)
 \right)\right|_{\epsilon=0}
 =\y_i^T  B_j(\z_i) \ .
\end{align*}
The directional derivative of the log-likelihood term, at the current estimate $\gamma$, is given by 
\begin{align}
b_j=n^{-1}\sum_{i=1}^n a_{ij}= n^{-1}\sum_{i=1}^n \y_i^T  B_j(  \gamma(\x_i)) \ .
\label{eq:likelihood_gradient}
\end{align}
We compute the directional derivative of roughness term numerically: for some small fixed $\epsilon>0$ set
$c_{j}
= \epsilon^{-1}(
R(\Psi_{ \epsilon  B_j}\circ  \gamma) -R( \gamma))$.
Thus, each coefficient $d_j$ is approximated as $d_j \approx b_j-\lambda c_j$, and the gradient is approximated by
$\nabla_{\gamma_{id}}H \approx \sum_{j=1}^L (b_j-\lambda c_j) B_j$.
Finally,  ${\gamma}$ is updated in the direction of $\nabla_{\gamma_{id}}H$ according to the mapping
\begin{align}
{\gamma} \mapsto \Psi_{(\delta \nabla_{\gamma_{id}}H)} \circ \gamma
\label{EQ:update_gamma}
\end{align}
for a small step size $\delta>0$.

{\bf Algorithm:}
\begin{enumerate}
\item
  Initialize ${\gamma}$ by the rigid rotation parametrized by $U_1U_2^T$,  where
   $D = U_1 \Sigma U_2^T$ is the modified singular value decomposition of 
   $D = \sum_{i=1}^n\y_i\x_i^T$ with  
   $U_1, U_2 \in \mathrm{SO}(2)$ and $\Sigma$ is the diagonal singular value matrix. 
\item
For each 
%
$j\in \{1,\ldots, L\}$, compute the log-likelihood coefficient ${b}_j$ 
according to equation (\ref{eq:likelihood_gradient}) and numerically compute the roughness coefficient ${c}_j$ for some small fixed step size $\epsilon>0$.
Use these coefficients to approximate the gradient $\nabla_{\gamma_{id}}H$.
\item
Update ${\gamma}$ for a small step size $\delta>0$ according to equation (\ref{EQ:update_gamma}).
\item
  If $E({\gamma})$ has converged, then stop. Otherwise return to step 2.

\end{enumerate}

\section{Experimental Results}\label{sec:experimets}
\subsection{Convergence Experiment Without Data}
\begin{figure}[h!]\centering
\begin{tabular}{ccc}
			\includegraphics[width=.3\textwidth]{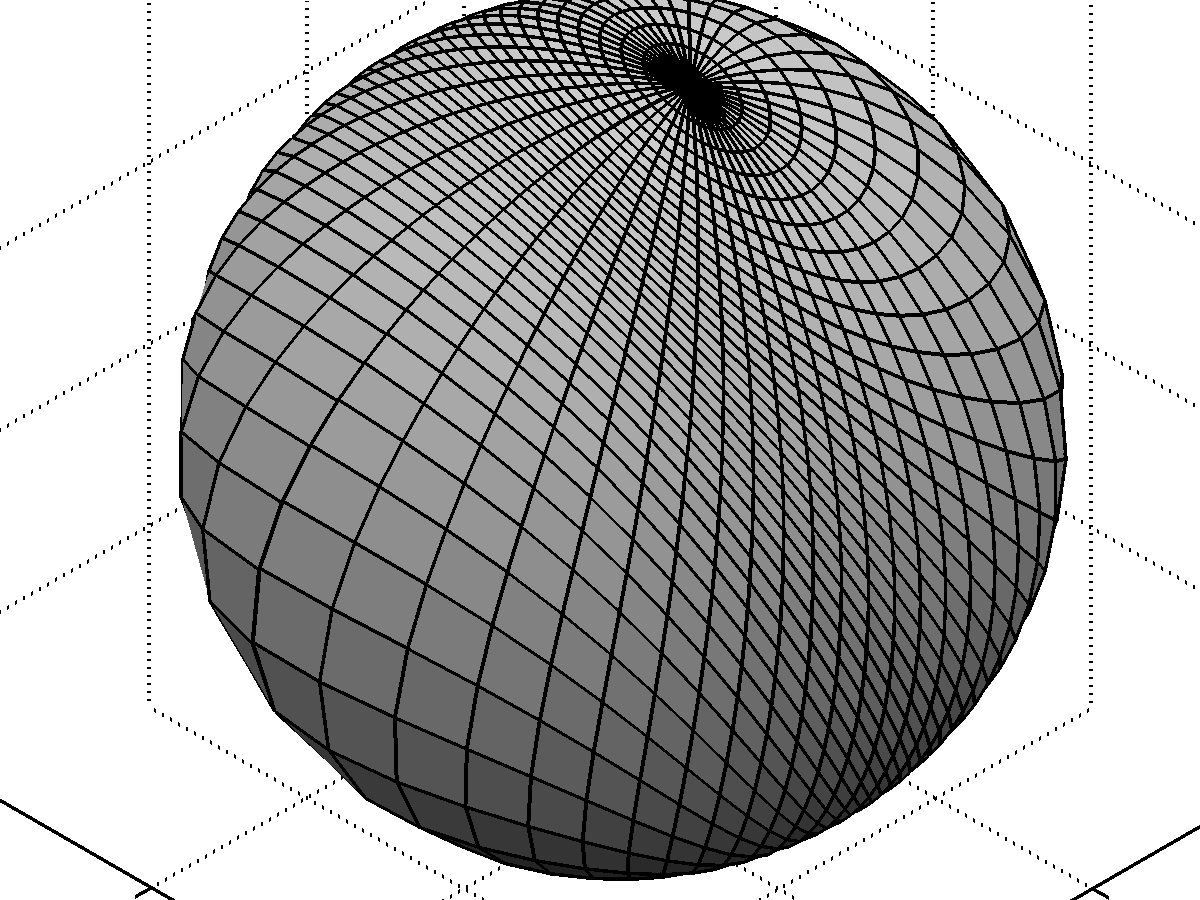}
			&
			\includegraphics[width=.3\textwidth]{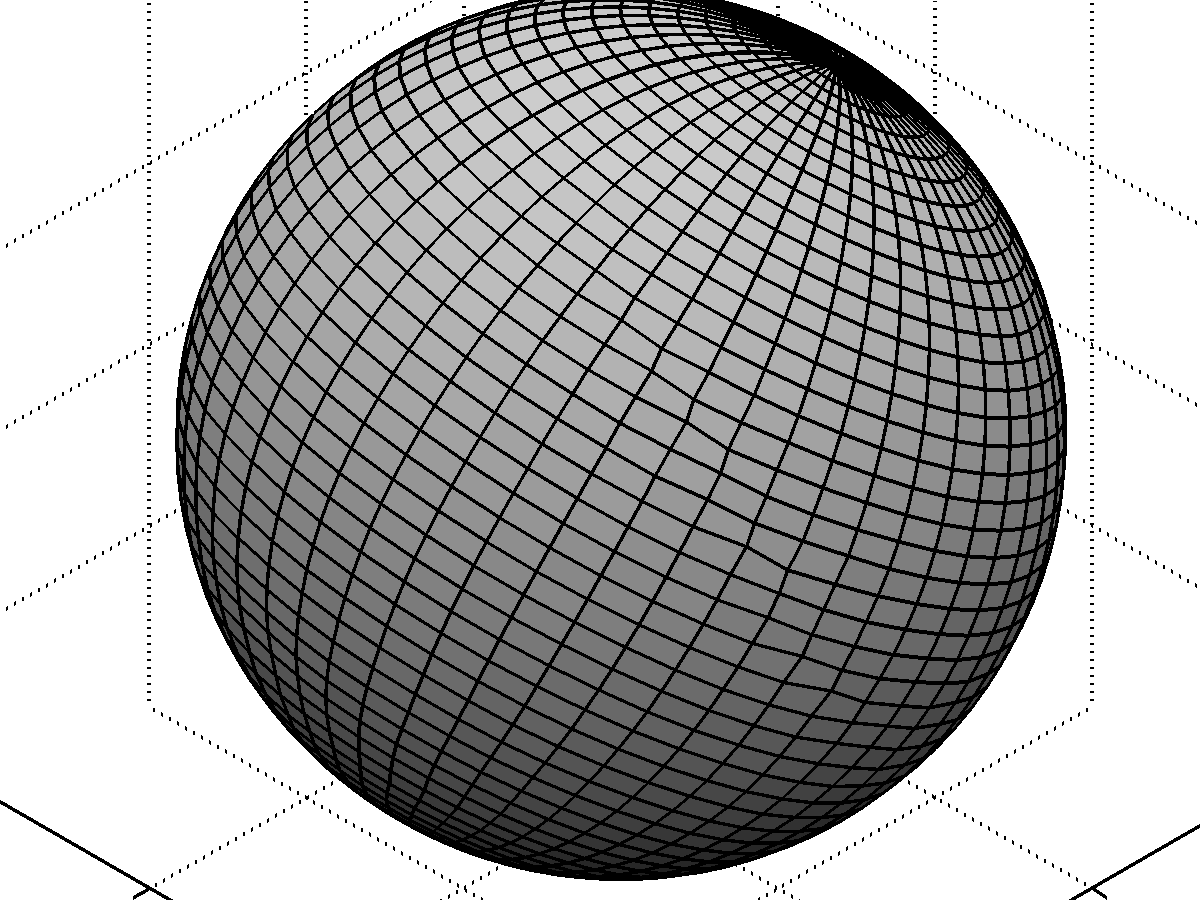}
			&
			\includegraphics[width=.3\textwidth]{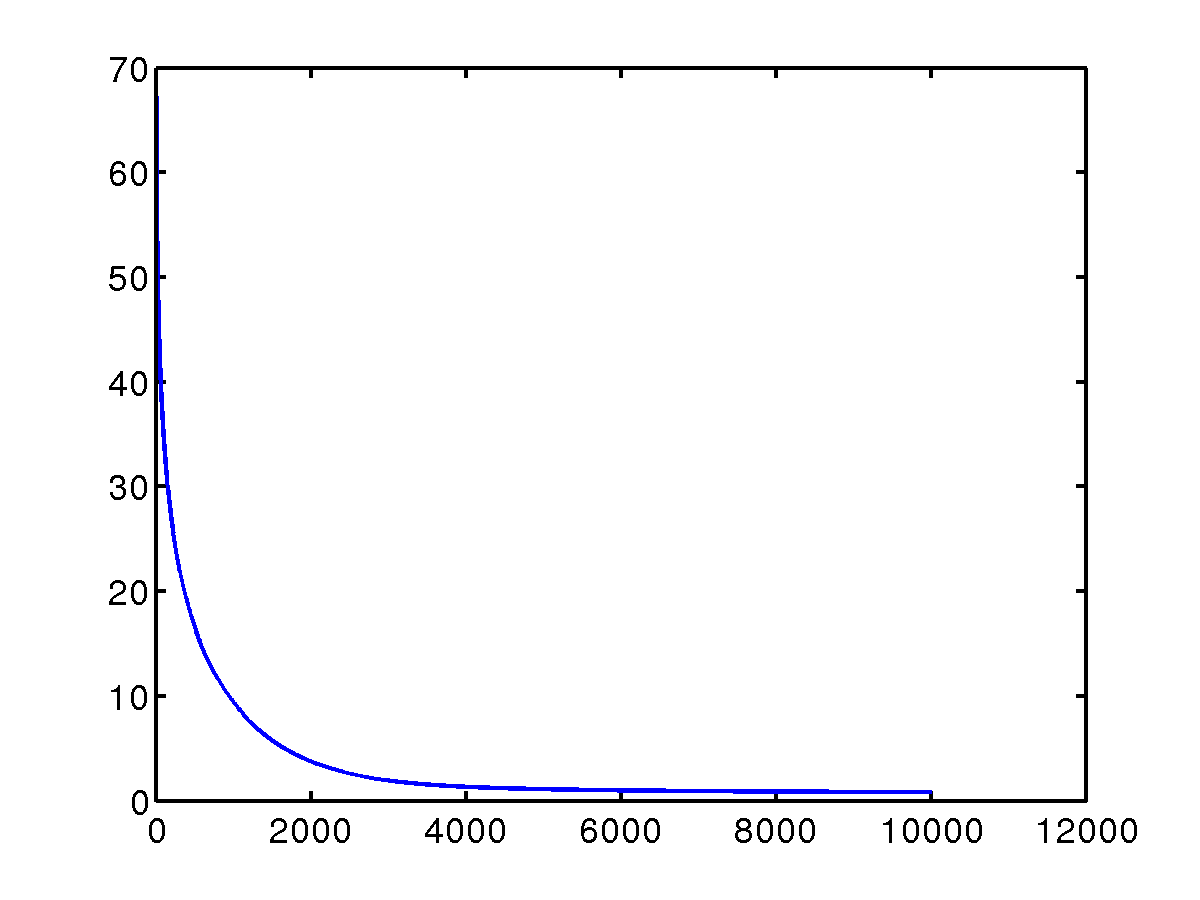}
\end{tabular}
\caption{
The left panel shows the initial diffeomorphism, the middle panel shows the resulting deformation after 10,000 iterations, and the right panel shows the evolution of the roughness measure.
}
\label{fig:convergence_experiment}
\end{figure}

In this experiment, we explore what happens with the gradient ascent algorithm when there is no likelihood term. We do this to check that the roughness term will push the deformation toward something which makes sense to have zero roughness. When there is no data, $n^{-1} \sum_{i=1}^n y_i^T\gamma(x_i)=0$ because the sum of the empty set is defined to be the additive identity. This implies that the objective function $E$ does not have a likelihood term in this case. We can perform this experiment using the previously define gradient and algorithm with one minor change.
Instead of initializing with a rigid rotation, we initialize our deformation with an arbitrary diffeomorphism and then iteratively apply the gradient for the roughness term. Our intuition is that by iteratively applying the gradient of the roughness term, the deformation will converge to something which closely resembles a rigid rotation. One can see in Fig. \ref{fig:convergence_experiment} that this initial deformation is relatively rough and distant from a rigid rotation. The resulting deformation after applying 10,000 iterations can be seen in the middle panel. The evolution of Roughness measure is plotted in the right panel. One can see further details and the animated results of this experiment in the supplementary material.

\subsection{Demonstration Using Simulated Data}
\begin{table}\begin{center}
\caption{Mean squared error of test data from Section 5 experiments.}
\begin{tabular}{cccccc} \\
\hline 
&OURS&NLL& PLT& RR&TRUE\\
\hline
Section 5.2&$18.9(10^{-2})$&$56.4(10^{-2})$&$22.2(10^{-2})$&$21.1(10^{-2})$&$1.82(10^{-2})$\\
Section 5.3&$4.1(10^{-3})$&$5.7(10^{-3})$&$19.1(10^{-3})$&$19.8(10^{-3})$&N/A\\
Section 5.4&$7.68(10^{-2})$&$11.6(10^{-2})$&$8.30(10^{-2})$&$9.42(10^{-2})$&N/A\\
\end{tabular}\\
\end{center}
OURS refers to the model presented in this paper. NLL refers to the nonparametric local linear regression model. PLT refers to the rigid rotation model. RR refers to the rigid rotation model. TRUE refers to the true diffeomorphism and is only applicable to the simulated experimental result from Section 5.2. N/A stands for Not Applicable.
\label{MSEtable}
\end{table}
We 
%
demonstrate the proposed nonparametric diffeomorphic regression model on both simulated and real data.
For the simulated case we compare the estimated model to the true deformation, and on the real data we compare predictive performances to some alternative models including Rigid Rotation 
%
%
from 
\cite{chang:1986}, Projective Linear Transformation 
%
%
from 
\cite{PLT:2012}, and the Nonparametric Local Linear 
%
model  
%
from 
 \cite{Taylor:2013}. We evaluate the predictive performance of a model by splitting a data set into training and test data. The model parameters 
are estimated using the training data, and then the performance is evaluated by computing the mean squared error of the test data. 
This error is computed as 
$MSE=n^{-1}\sum_{i=1}^n \|\y_i- \hat{{\mu}}_{i}\|^2 $, where $\y_i$ and  $\hat{{\mu}}_{i}$, 
respectively, denote the true and predicted values and $\|\cdot \|$ denotes the Frobenius norm. 
Since 
the data and the predicted values are restricted to a sphere, the mean squared error lies in [0, 4]. We also compare the fitted spherical mappings of each model and compare the observed and predicted values.

We start by illustrating the gradient ascent algorithm using simulated paired data points on $\s^2$ that are related by a diffeomorphism $\gamma_0\in \Gamma$. The true $\gamma_0$ is obtained via a combination of M\"obius and twisting
transformations
on $\s^2$.  Additionally, we compose a sequence of small incremental diffeomorphisms using spherical harmonics to obtain our final true diffeomorphism. See the supplementary for details on these maps and their parametrization for this example.
Once $\gamma_0$ is generated, it is fixed throughout the experiment. 
The training and test data respectively consist of $200$ and $100$  independently 
sampled
data points.
Predictor variables $\x_1,\ldots,\x_n$ are  simulated independently from a von-Mises Fisher distribution with mean $(0,0,1)^T$ and concentration $\kappa=5$ for the training data and uniformly for the test data. This will leave a gap in the training data to simulate model performance with limited information. The individual responses $\y_i$ are then independently simulated from a von-Mises Fisher distribution with mean $ \gamma_0(\x_i)$ and concentration $\kappa=100$. The training and test data can be seen in the supplementary. 

\begin{figure}[h!]\centering
\begin{tabular}{cccc}
True Diffeomorphism  &$\lambda=(10^{-10})$&$\lambda=2.5(10)^{-4}$&$\lambda= (10)^{-3}$\\
&$l=3$&$l=3$&$l=3$\\
\includegraphics[width=.22\textwidth]{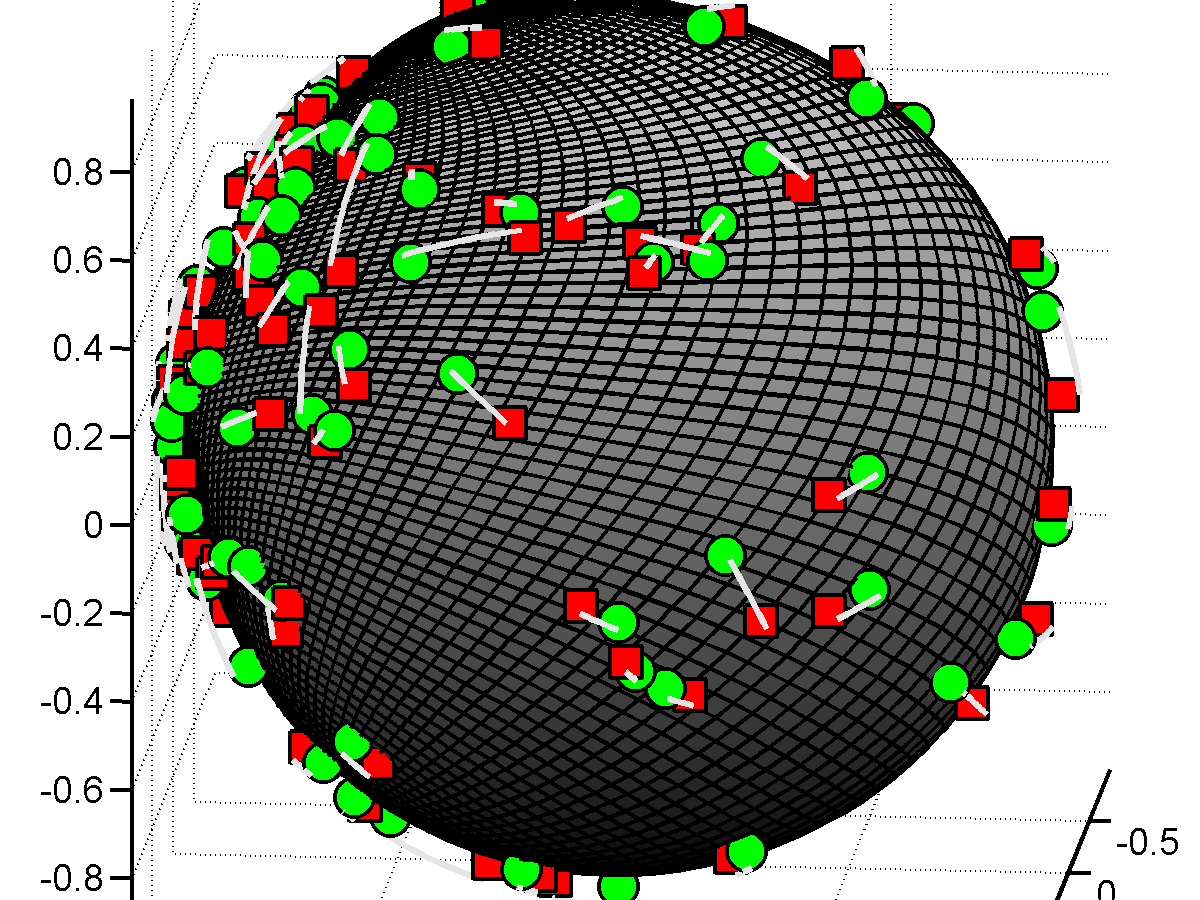}&
\includegraphics[width=.22\textwidth]{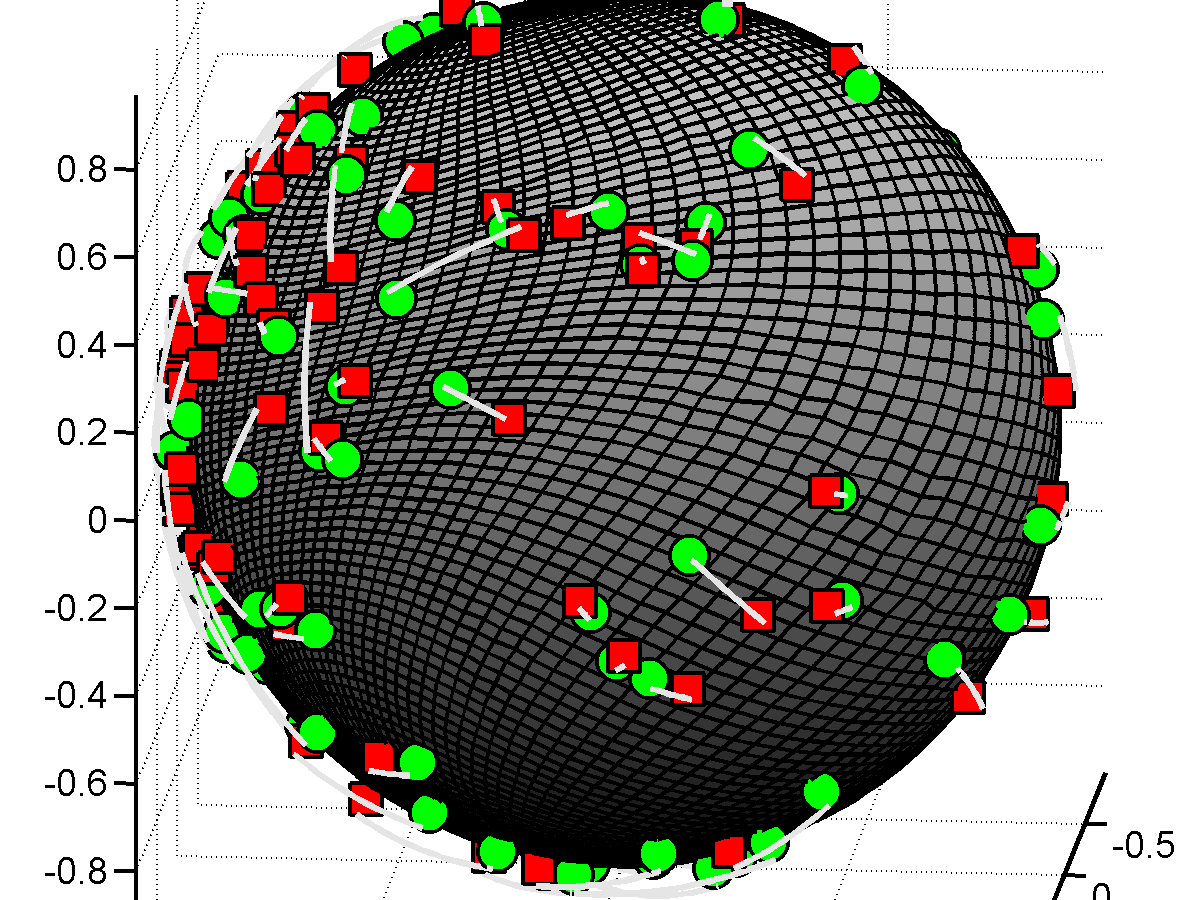}&
\includegraphics[width=.22\textwidth]{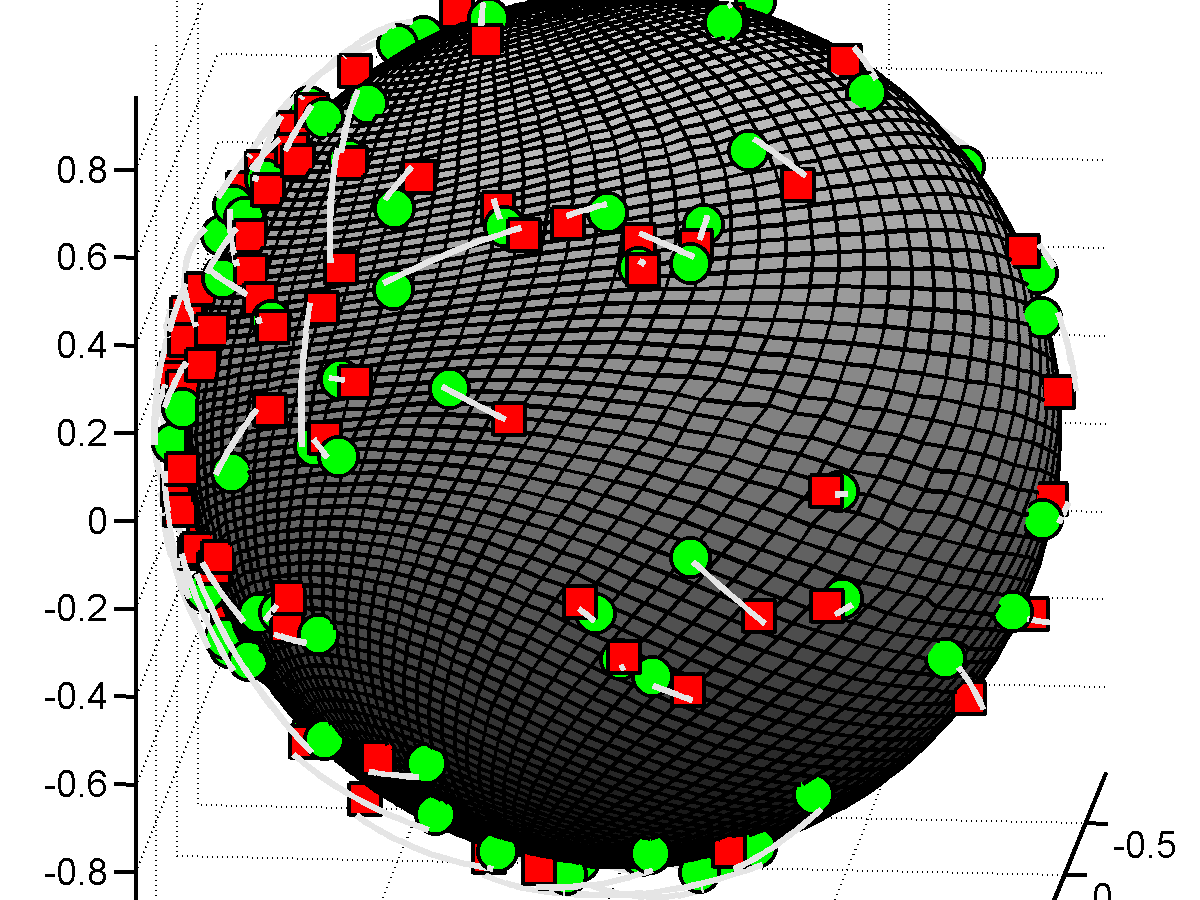}&
\includegraphics[width=.22\textwidth]{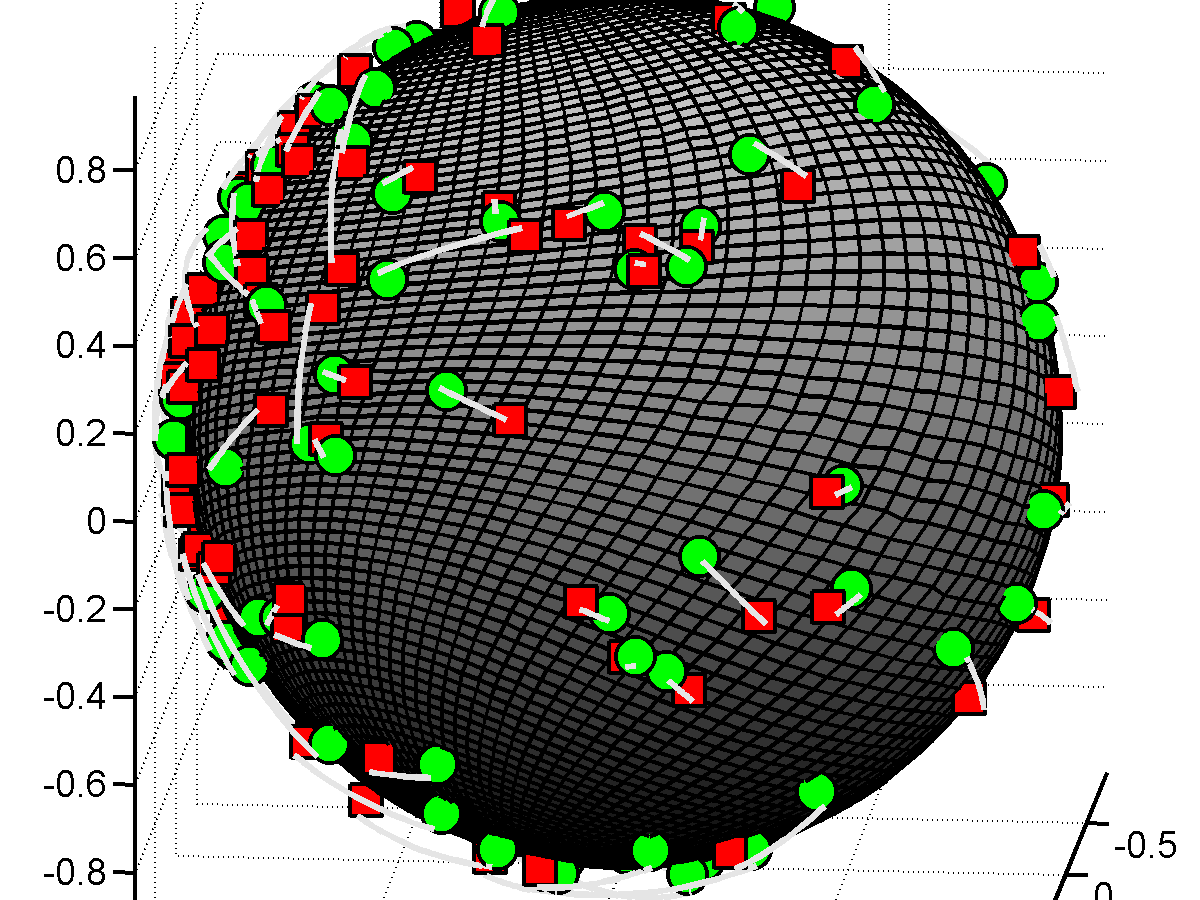}\\\hline
&$\lambda= (10^{-4})$&$\lambda=2.5(10)^{-4}$&$\lambda= 7.5(10)^{-4}$\\
&$l=10$&$l=10$&$l=10$\\
&
\includegraphics[width=.22\textwidth]{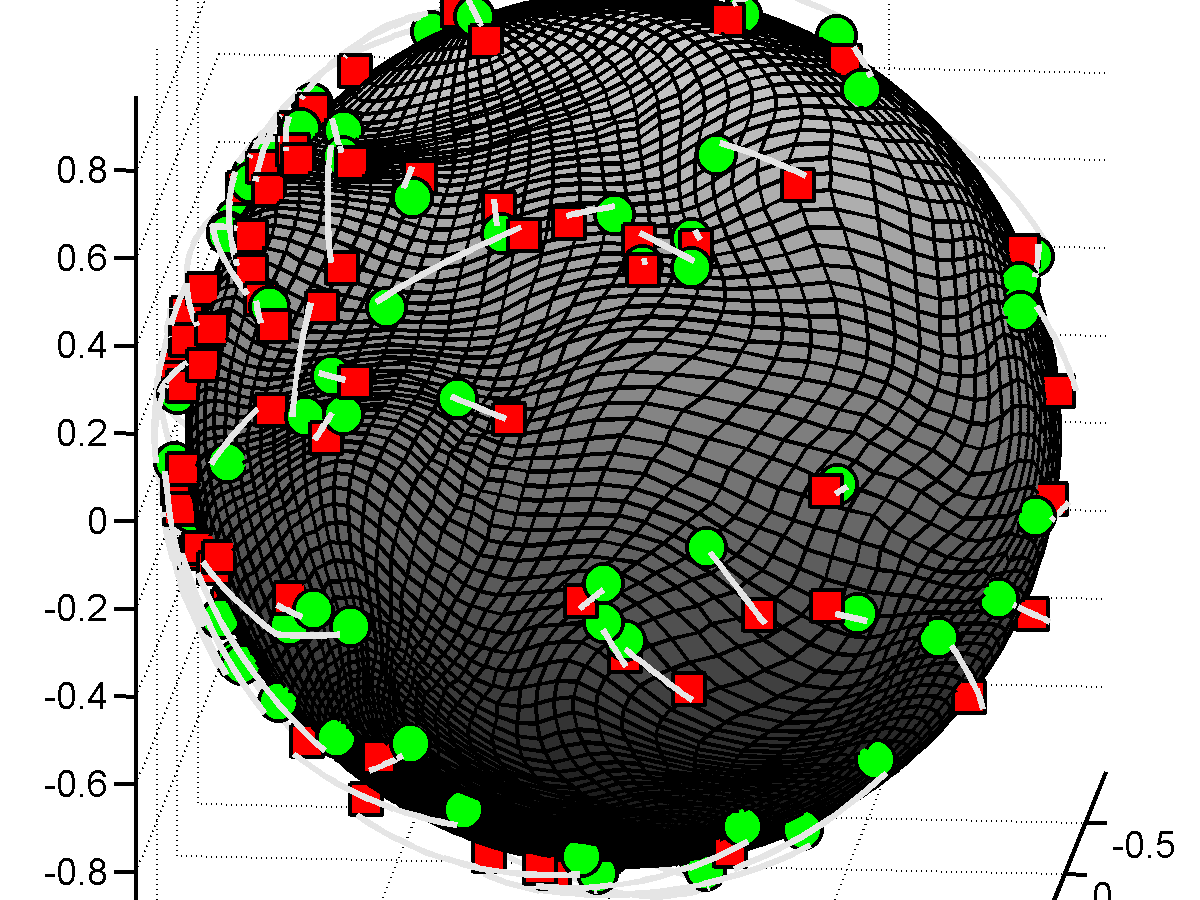}&
\includegraphics[width=.22\textwidth]{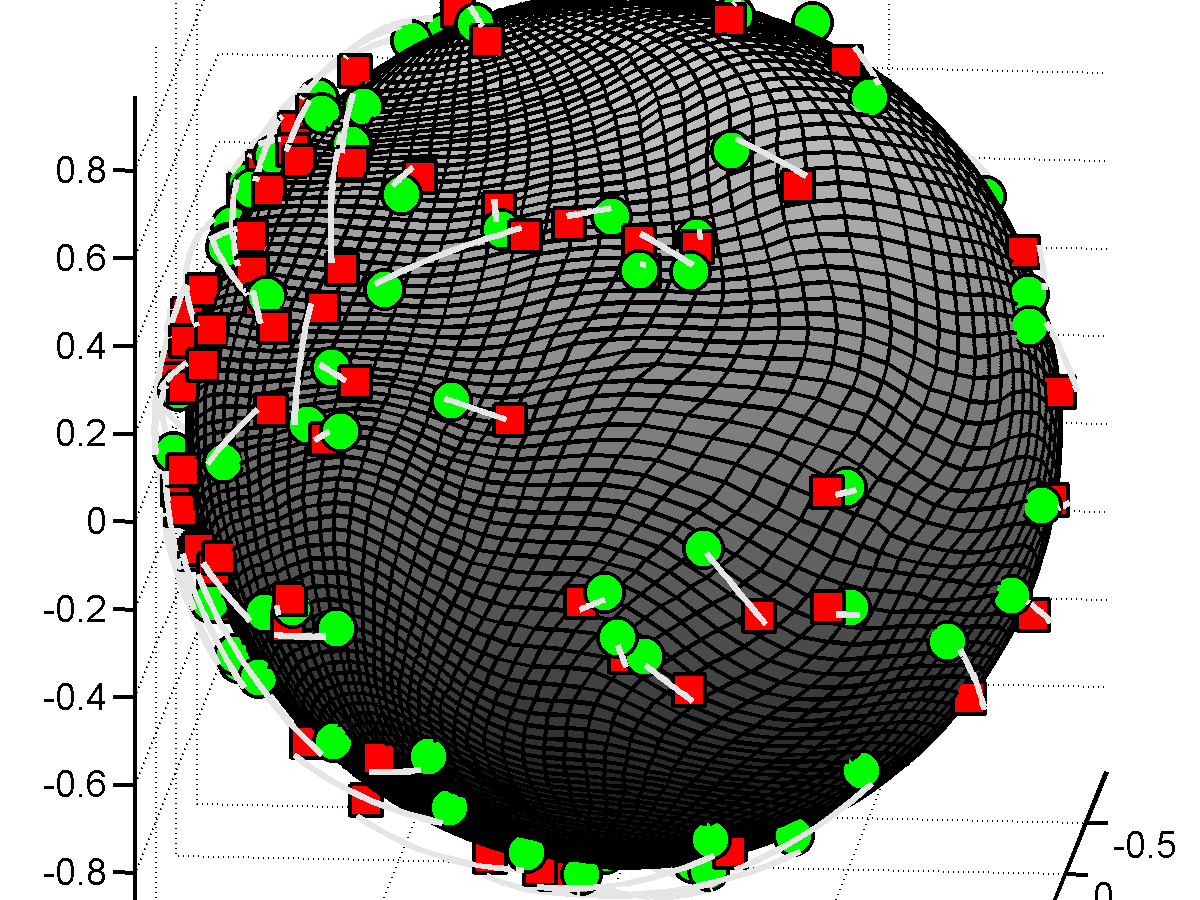}&
\includegraphics[width=.22\textwidth]{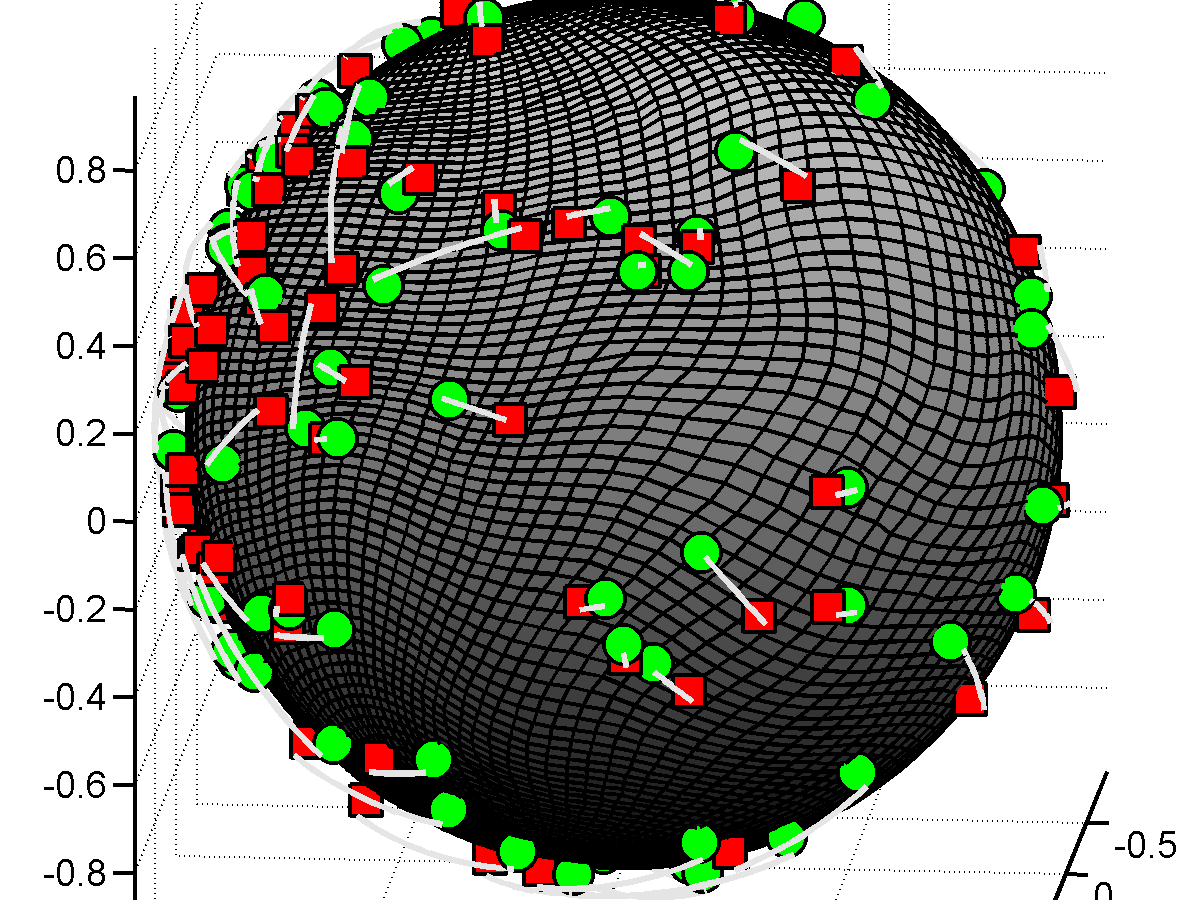}\\
\end{tabular}
\caption{
This illustration compares the observed (in red) and predicted (in green) test values of the true diffeomorphism and deformations fitted at various $\lambda$ values.
}
\label{fig:simdiffeo_lambda}
\end{figure}

The roughness parameter $\lambda$ is estimated using cross-validation, i.e. by splitting the training data into 75 training observations and 25 validation observations. 
In the left panel of Fig. \ref{fig:simdata}, one can see the mean squared error of the validation data plotted over various levels of $\lambda$ and for $l\in \{3,10\}$. We select the value of $\lambda$ and $l$ which minimizes the validation error. In this case, we construct a basis from the spherical harmonic functions of order less than or equal to $l=3$, select $\lambda=2(10)^{-4}$, and use the model fitted using these values to make future predictions.  
The right panel of Fig. \ref{fig:simdata} shows the evolution of $E({\gamma})$ as ${\gamma}$ is iteratively updated according to the gradient ascent algorithm applied to the full set of $200$ training data points.
In Fig. \ref{fig:simdiffeo_lambda}, 
the true diffeomorphism $\gamma_0$ is compared to estimated diffeomorphisms at several roughness levels.  Notice for $l=3$ that the difference in the estimated diffeomorphisms are very subtle over $\lambda$. The roughness penalty's influence is more visible for $l=10$.

A comparison of other model performances on the test data is summarized in the first row of Table \ref{MSEtable}. None of the estimated models outperform the true diffeomorphism because data is scarce over an part of the sphere. The nonparametric local linear regression model performs poorly here because it does not handle the extrapolation well. All the other models outperform the local linear model because they assume the underlying deformation is a diffeomorphism, which is true in this simulated the case. One can see a comparison of estimated deformations for other models with observed and predicted values plotted in the supplementary material.
\begin{figure}[h!]\centering
\begin{minipage}[c]{.4\textwidth}\begin{center}
$MSE$\\
			\includegraphics[width=\textwidth]{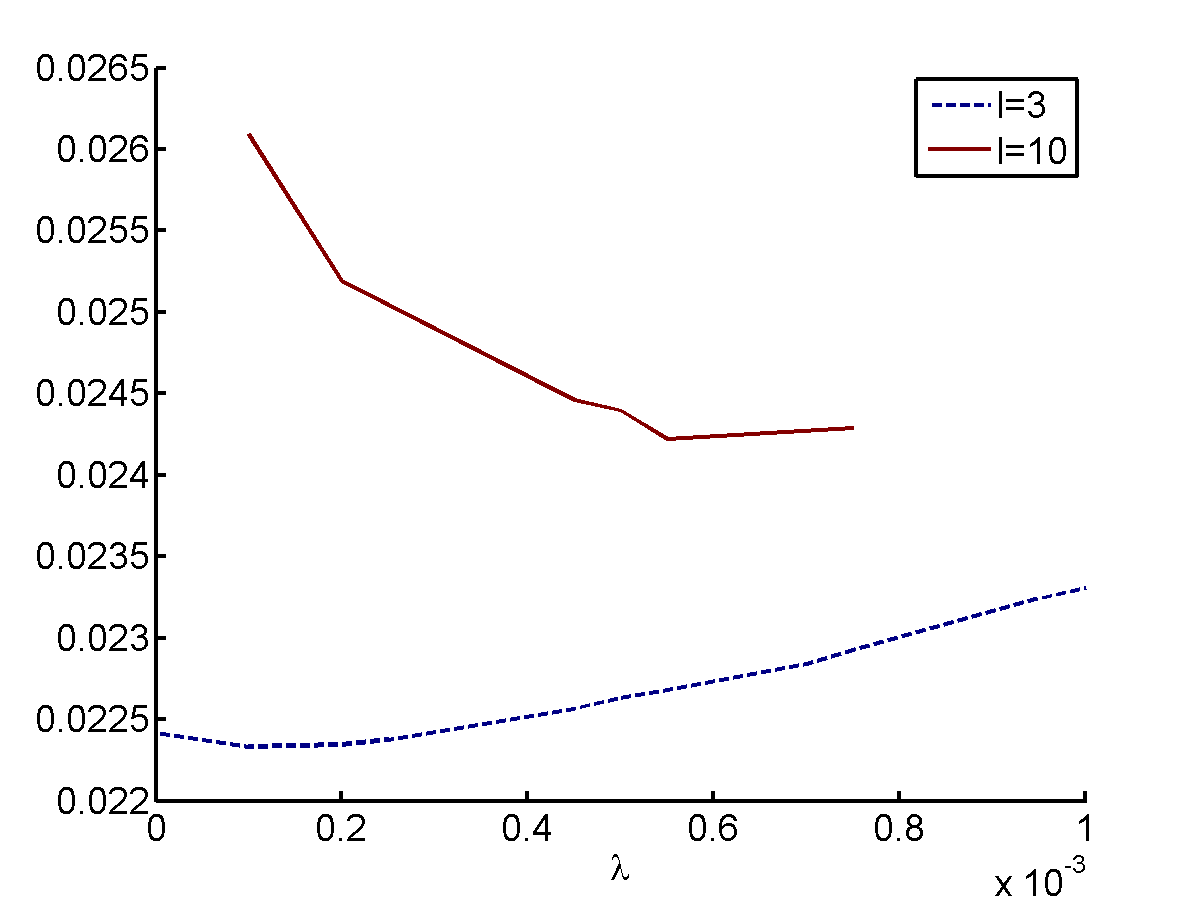}
			$\lambda$
			\end{center}
\end{minipage}
\begin{minipage}[c]{.4\textwidth}\begin{center}
$E({\gamma})$
\\  \vspace{6pt}
			\includegraphics[width=\textwidth]{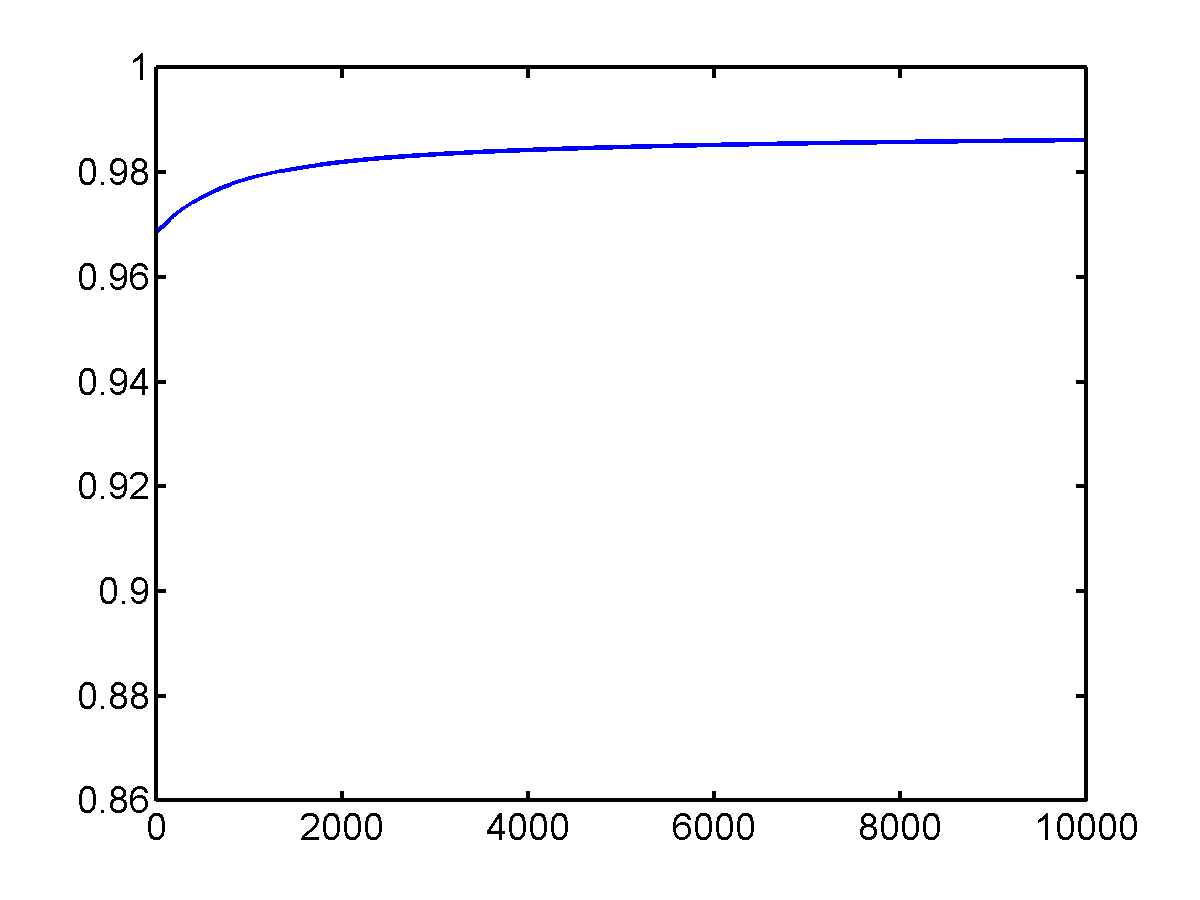}\\
Iteration
			\end{center}
\end{minipage}
\caption{
On the left, one can see the mean squared error on the validation data for various values of $\lambda$. The dotted blue and solid red line respectively denote $l=3$ and $l=10$.
The right plot show the evolution of the penalized log-likelihood function as the algorithm iterates. 
}
\label{fig:simdata}
\end{figure}

\subsection{Weather Balloon Wind Directions Data}
\begin{figure}[h!]\centering
\begin{tabular}{ccc}
Training Data ($n=200$)&Test Data ($n=494$)&\\
\begin{minipage}[c]{.33\textwidth}
\includegraphics[width=\textwidth]{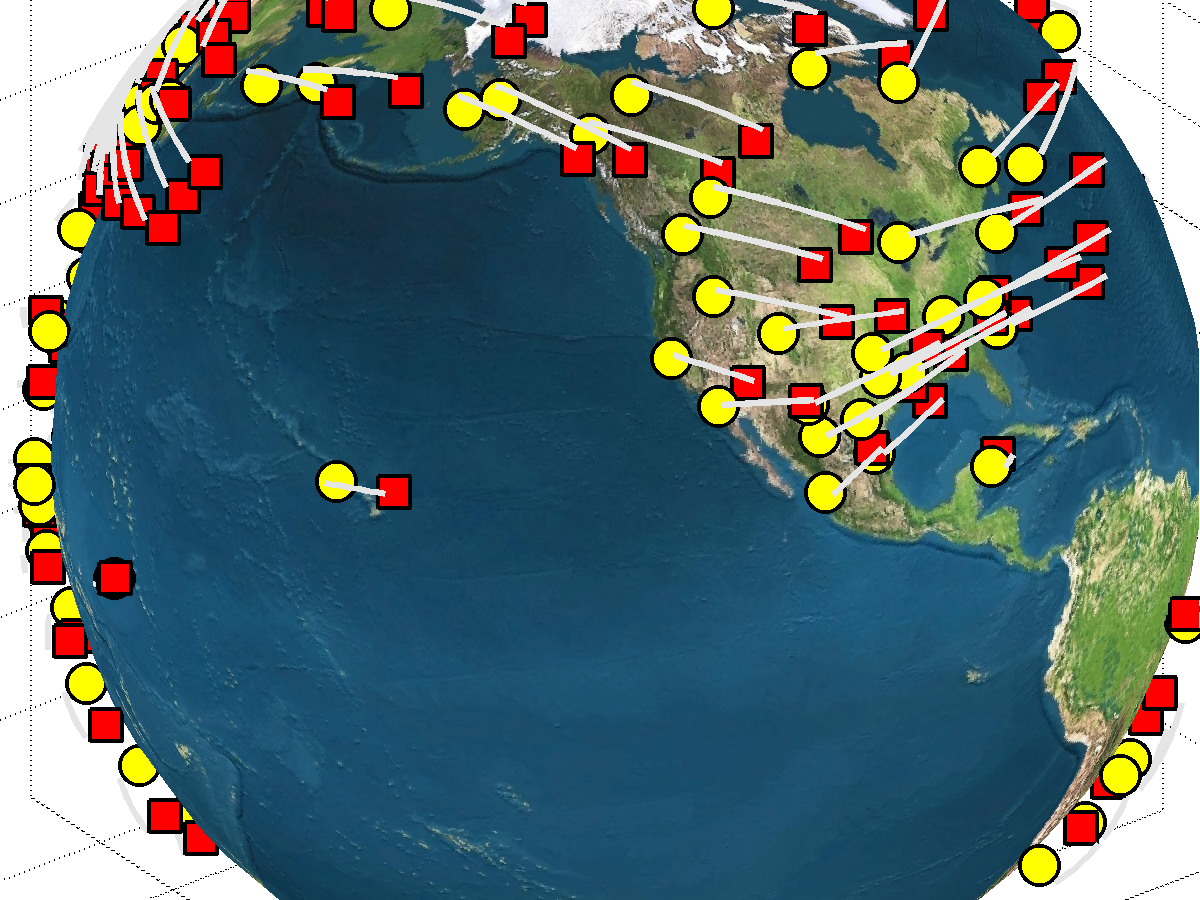}
\end{minipage}
&
\begin{minipage}[c]{.33\textwidth}
\includegraphics[width=\textwidth]{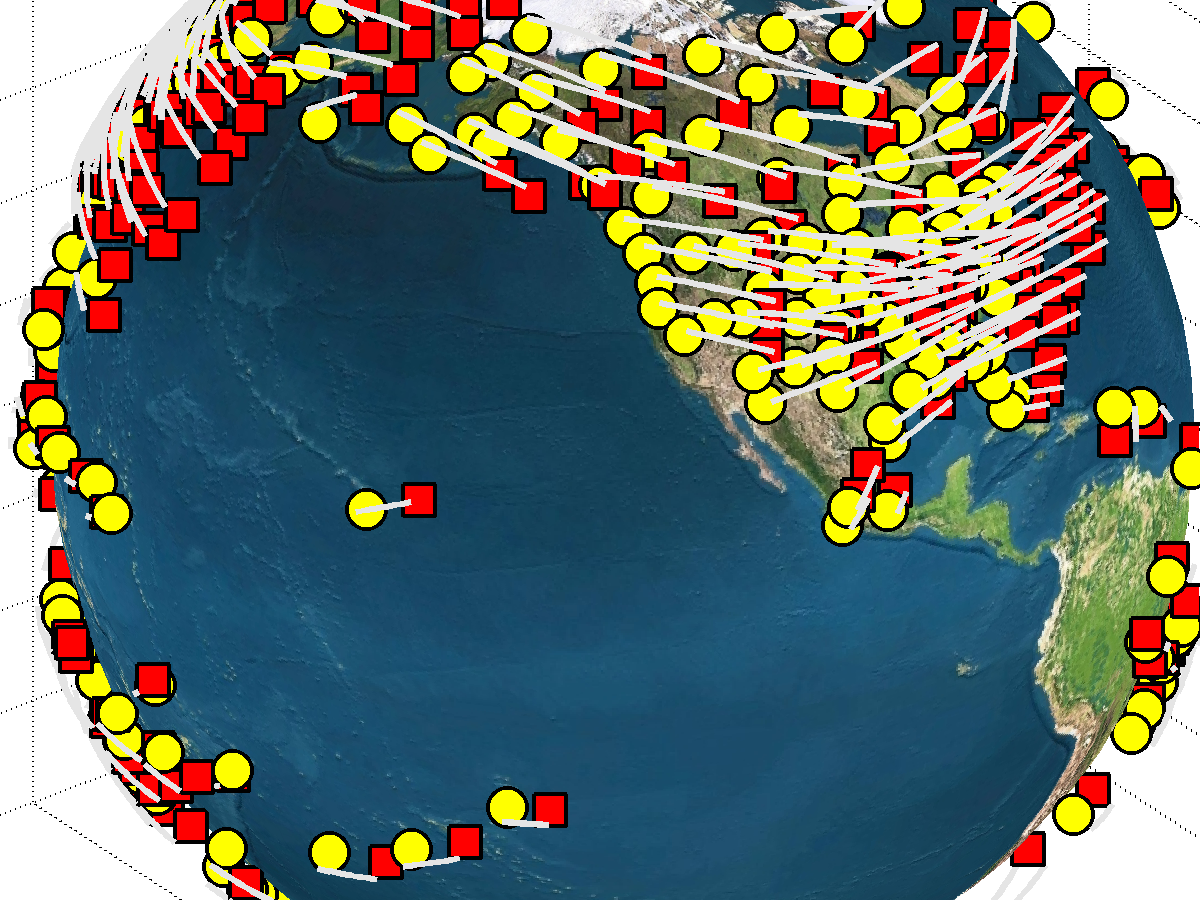}
\end{minipage}
&
\end{tabular}
\caption{ The figures show the training and test data for the wind direction models. The yellow dots denote the locations of the weather monitoring stations. The red dots denote the average wind displacement. Corresponding data points are connected with a light gray lines. 
}
\label{figure:BalloonTrainTest}
\end{figure}

The Integrated Global Radiosonde Archive contains radiosonde and pilot balloon observation from various stations distributed across the globe. An overview of data coverage 
is presented in \cite{Durre:2006}. The data set was constructed from the monthly average of November 2013. In this data set, there are typically multiple means per station which correspond to various pressure levels. We take an average of tangential wind velocities at all the available pressure levels at a point to form a direction at that point. This results in 694 spatial observations, which we split into 200 training observations and 494 test observations. For each observed $i=1,\ldots, n$ the station location will be treated as the predictor $\x_i$, and the corresponding tangential wind velocities will be denoted by $\v_i \in T_{\x_i}(\mathbb{S}^2)$. 
The tangential wind velocities are measured in scaled units of 
$600$ meters per second
for this data. 
Each
response variable $\y_i = \exp_{\x_i}(\v_i)$ 
is
obtained by applying the exponential map to each respective tangent vector. The training data, test data are shown in Fig. \ref{figure:BalloonTrainTest}. 
Using cross validation on the training data, we selected $\lambda=(10)^{-10}$ for the roughness penalty of our model and $l=10$ for the maximum order of basis elements. A plot of the $MSE$ for the validation data used for tuning the model is in the supplementary.

\begin{figure}[h!]\centering
\begin{tabular}{cc}
OURS&NLL\\
\includegraphics[width=.4\textwidth]{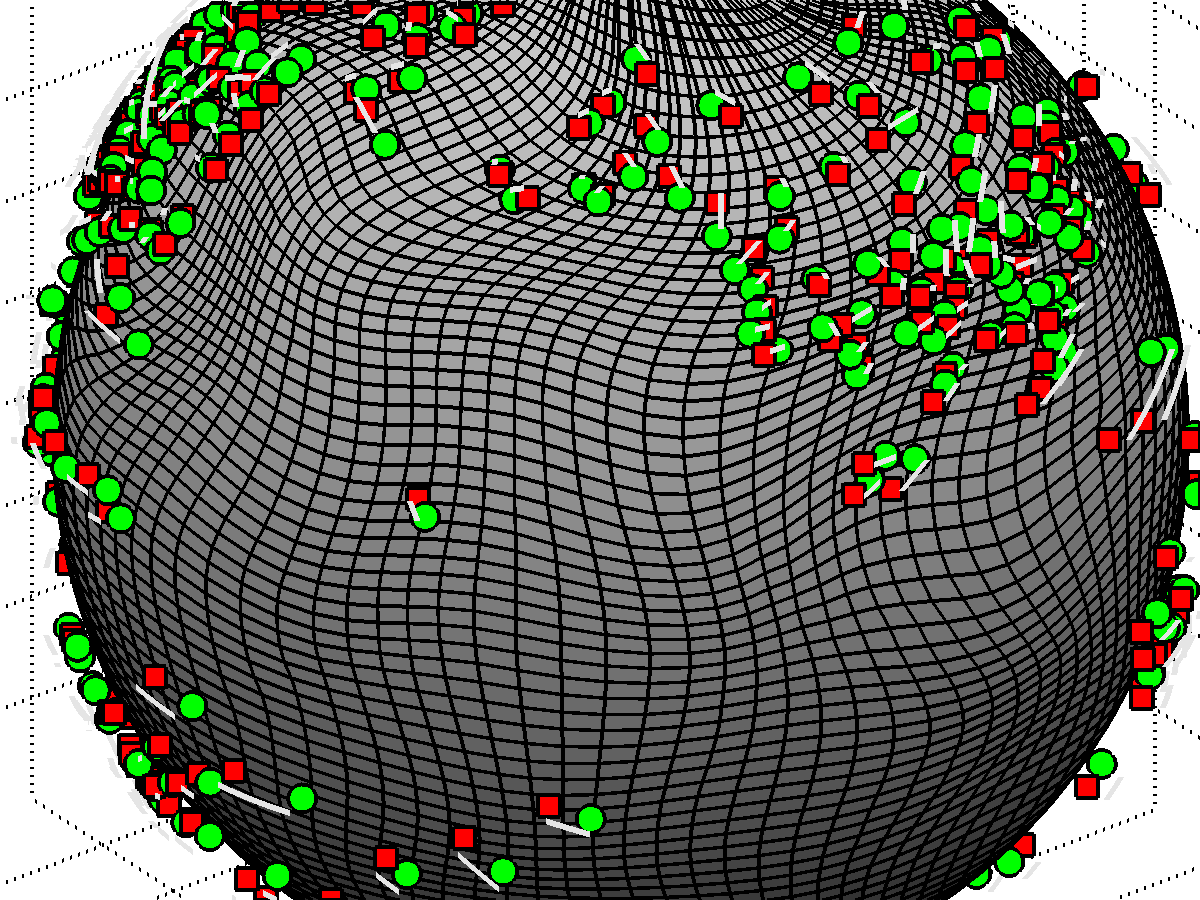}
&
\includegraphics[width=.4\textwidth]{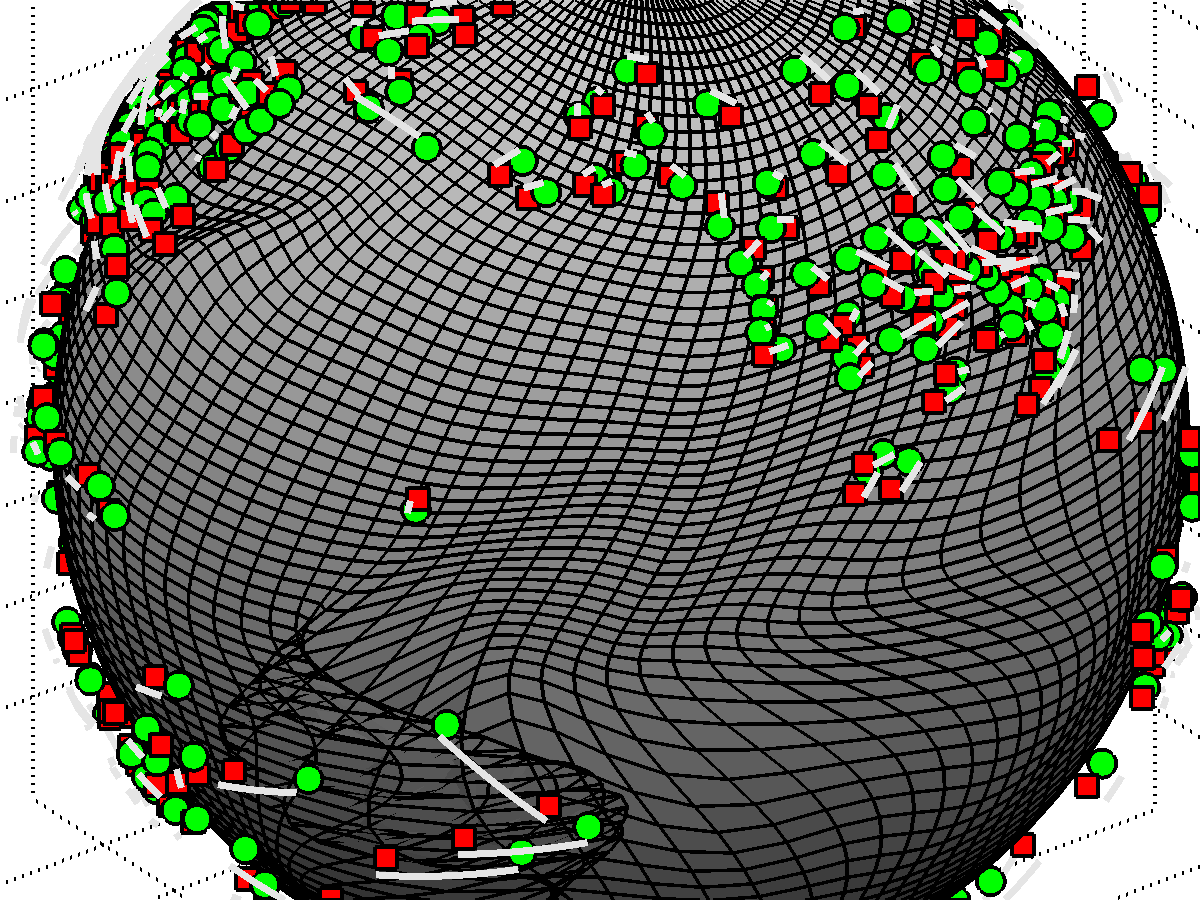}
\\
PLT&RR\\
\includegraphics[width=.4\textwidth]{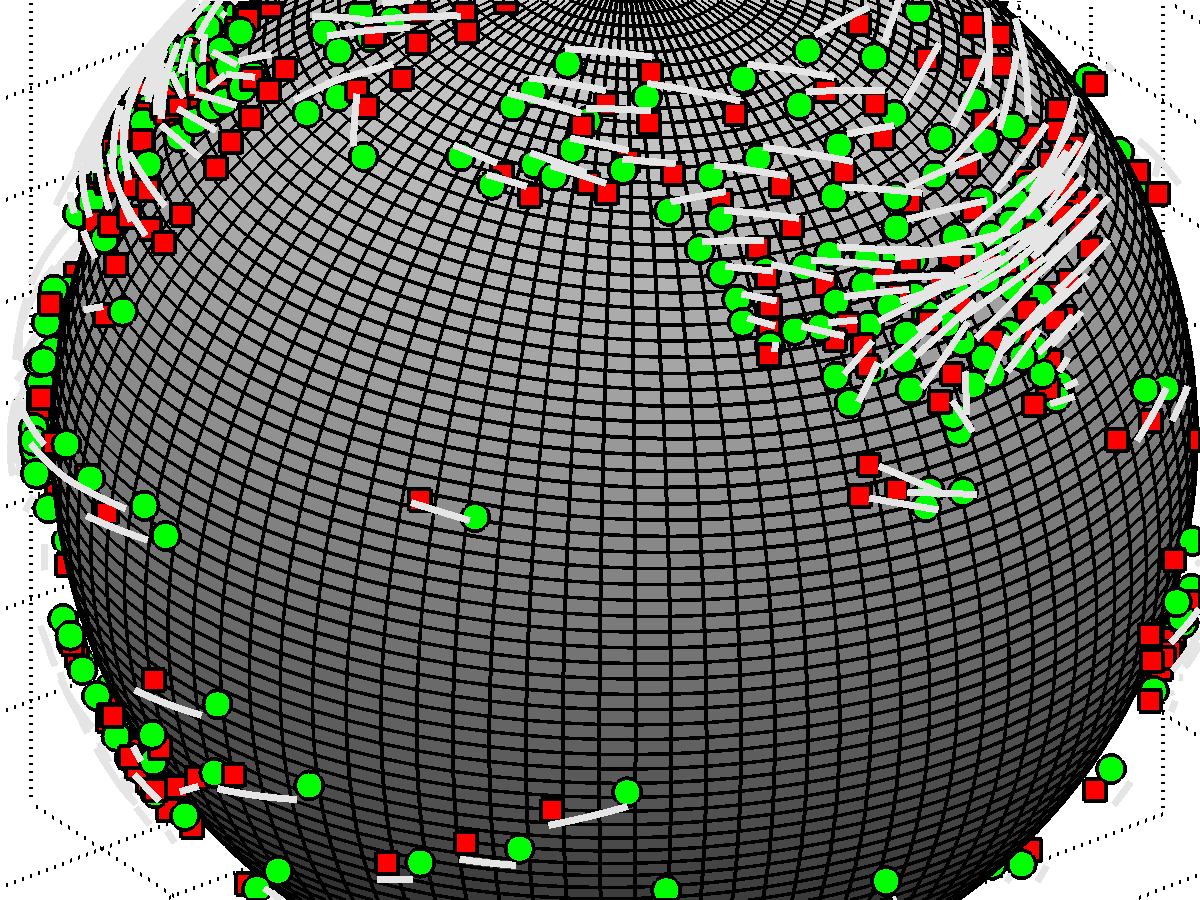}
&
\includegraphics[width=.4\textwidth]{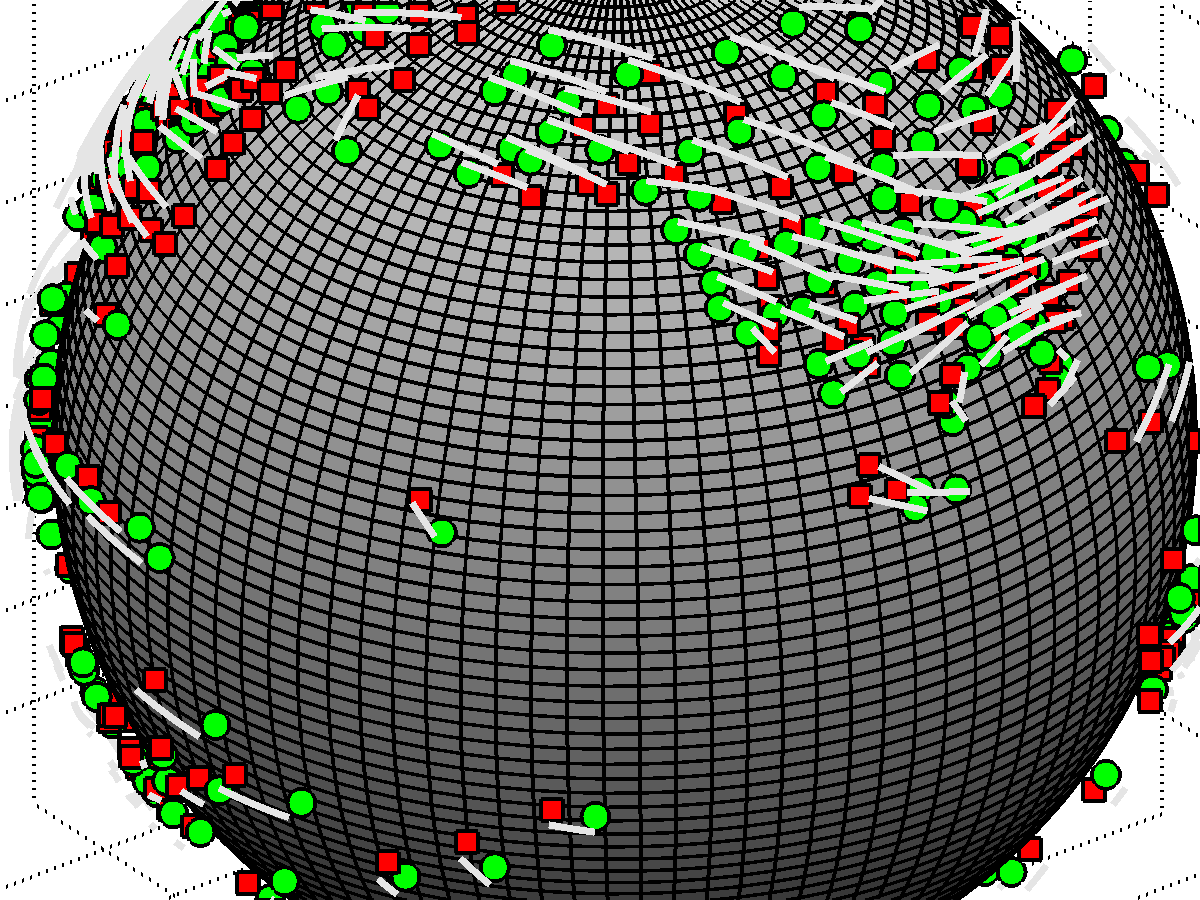}\\
\end{tabular}
\caption{Deformations of fitted models from weather training data with observed and predicted test data respectively plotted as red squares and green circles. Corresponding data points are connected with a light gray lines. OURS refers to the model presented in this paper. NLL refers to the nonparametric local linear regression model. PLT refers to the rigid rotation model. RR refers to the rigid rotation model.}
\label{fig:balloon}\end{figure}
 The test error in Table \ref{MSEtable} shows that out method outperforms the nonparametric local linear model. In Fig. \ref{fig:balloon}, the observed and predicted tangential wind directions are plotted for each model. The nonparametric local linear model is in close agreement with our diffeomorphic model in regions where there is abundant data available. In places where training data is scarce, as seen in the south pacific region, the nonparametric
local linear model is non-injective and allows the mesh grid to overlap and deform heavily.

\subsection{Vector-Cardiogram Data}
A vector-cardiogram measures the direction and magnitude of electrical forces that are generated by heart actions. 
The directional aspect of these vectors have important applications in the diagnoses of certain diseases. The dataset, which was used in \cite{downs:2003}, 
consists of vector-cardiogram data from 98 children ages 2-18, where each child is measured using two 
lead
systems, namely the Frank system and the McFee system. The objective of this experiment is to define a correspondence between these two systems.  To do this, models will be fitted using the directional vector from the 
Frank system as the predictor and the directional vector from the Mcfee system as the response. 
This is could be useful for combining data sets that use two different systems. In this case, we choose to convert Frank system to optimally correspond with Mcfee system data.

\begin{figure}[h!]\centering
\begin{tabular}{ccc}
Training Data ($n=70$)&Test Data ($n=28$)&\\
\begin{minipage}[c]{.33\textwidth}
\includegraphics[width=\textwidth]{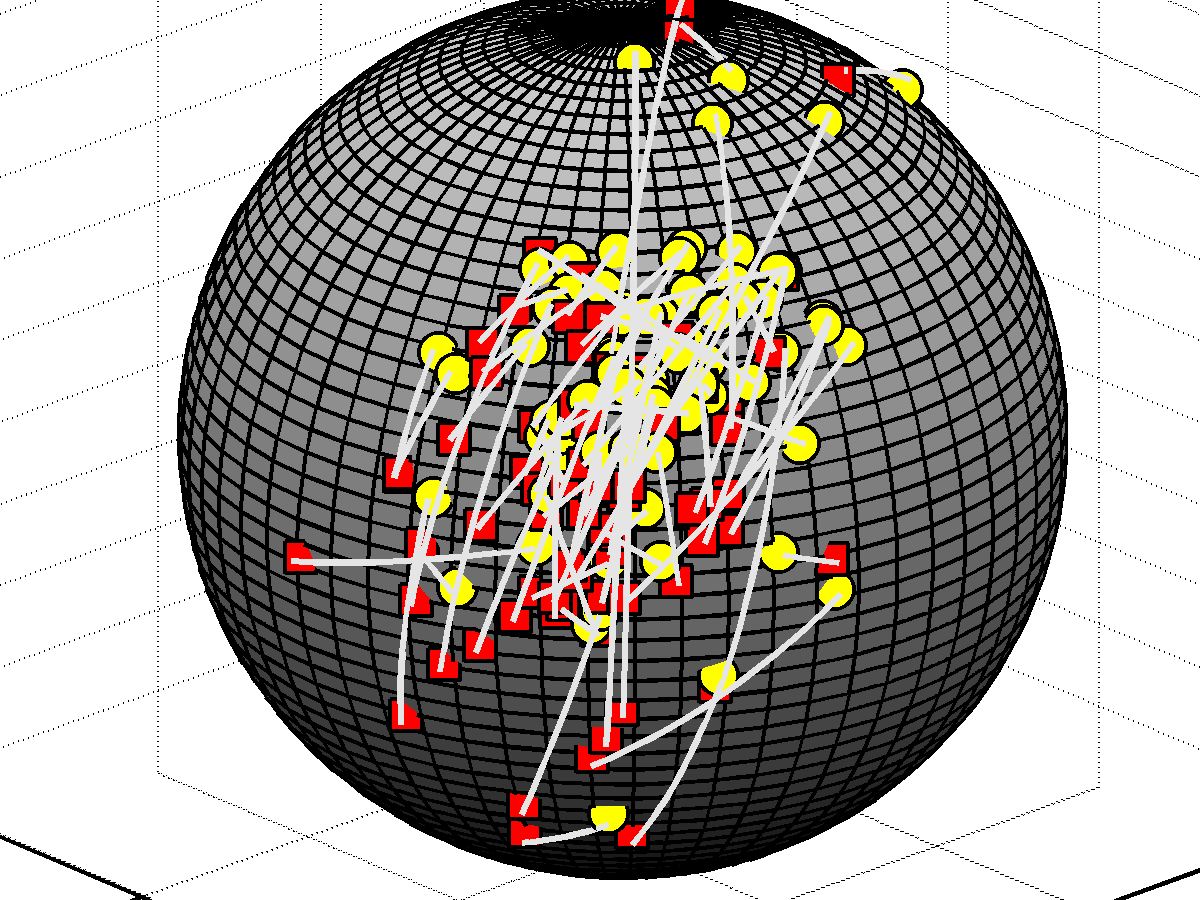}
\end{minipage}
&
\begin{minipage}[c]{.33\textwidth}
\includegraphics[width=\textwidth]{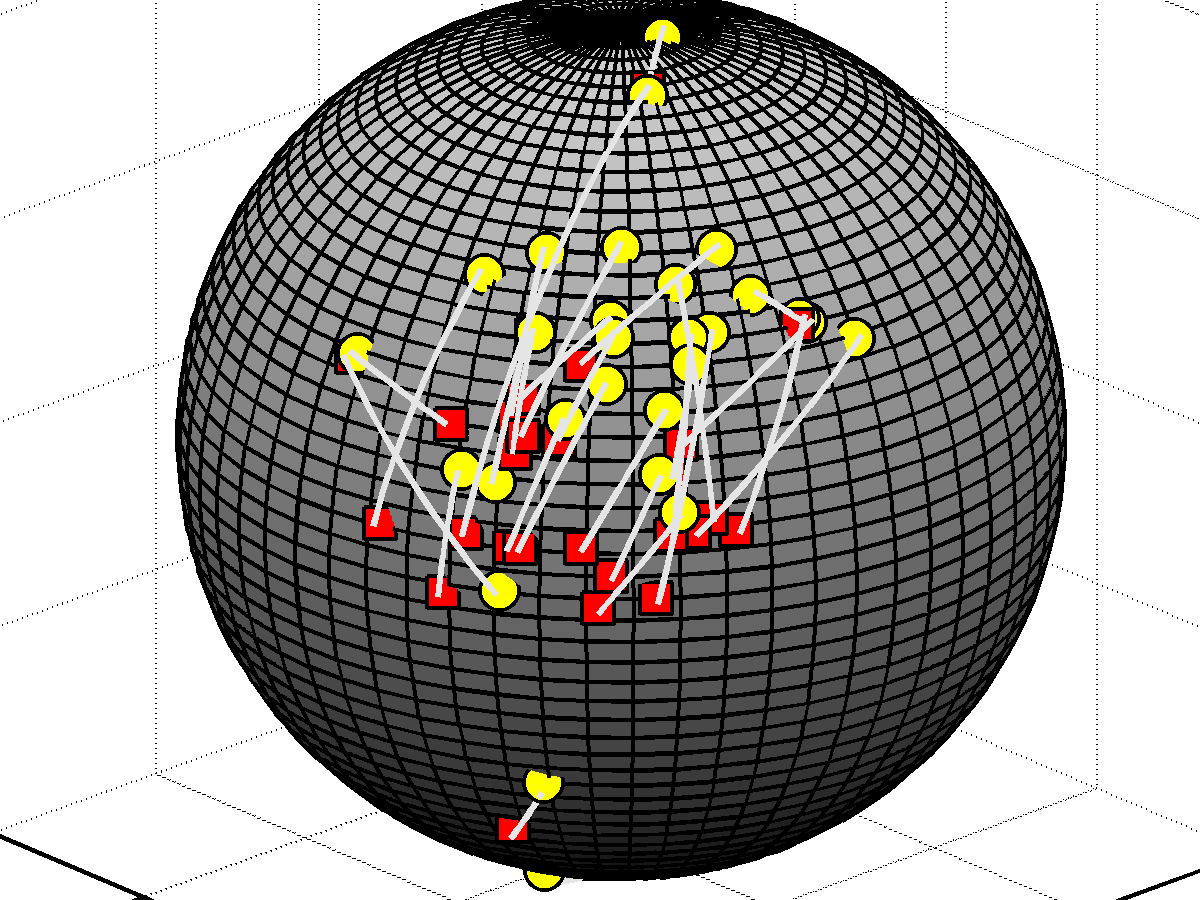}
\end{minipage}
&
\end{tabular}
\caption{ The figure shows the direction of greatest magnitude using the Frank system in yellow and the McFee system in red. Corresponding data points are connected with a light gray lines. 
}
\label{figure:vector cardiogram}
\end{figure}

\begin{figure}[h!]\centering
\begin{tabular}{cc}
Ours&NLL\\
\includegraphics[width=.4\textwidth]{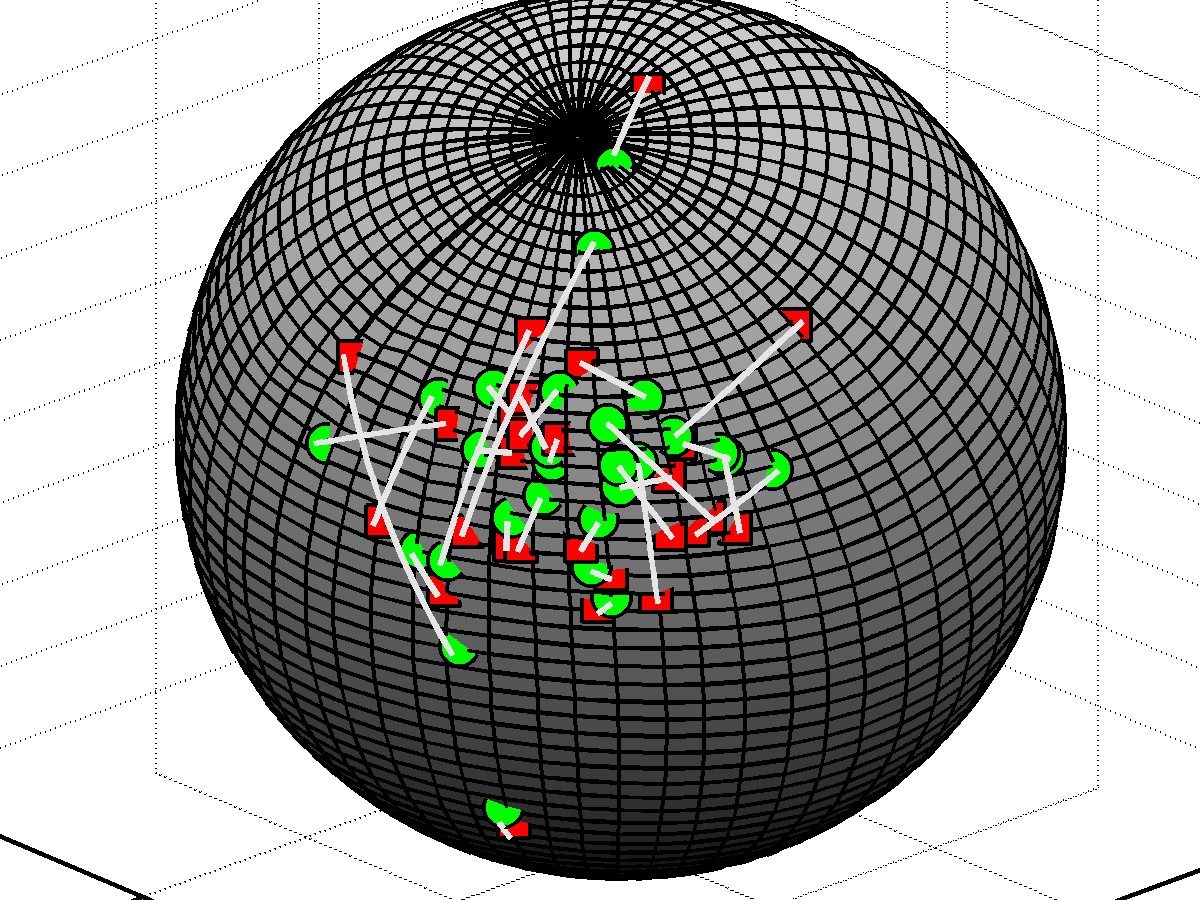}
&
\includegraphics[width=.4\textwidth]{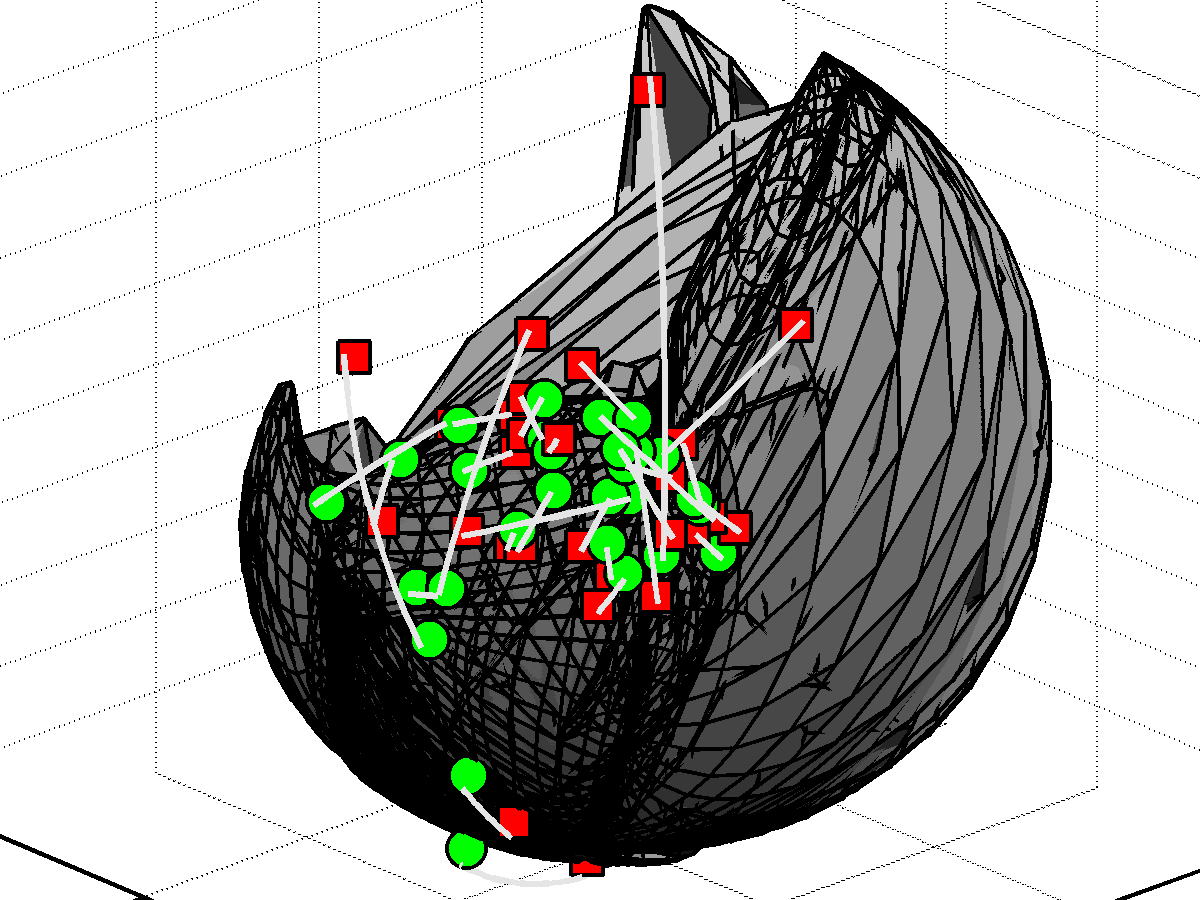}
\\
PLT&RR\\
\includegraphics[width=.4\textwidth]{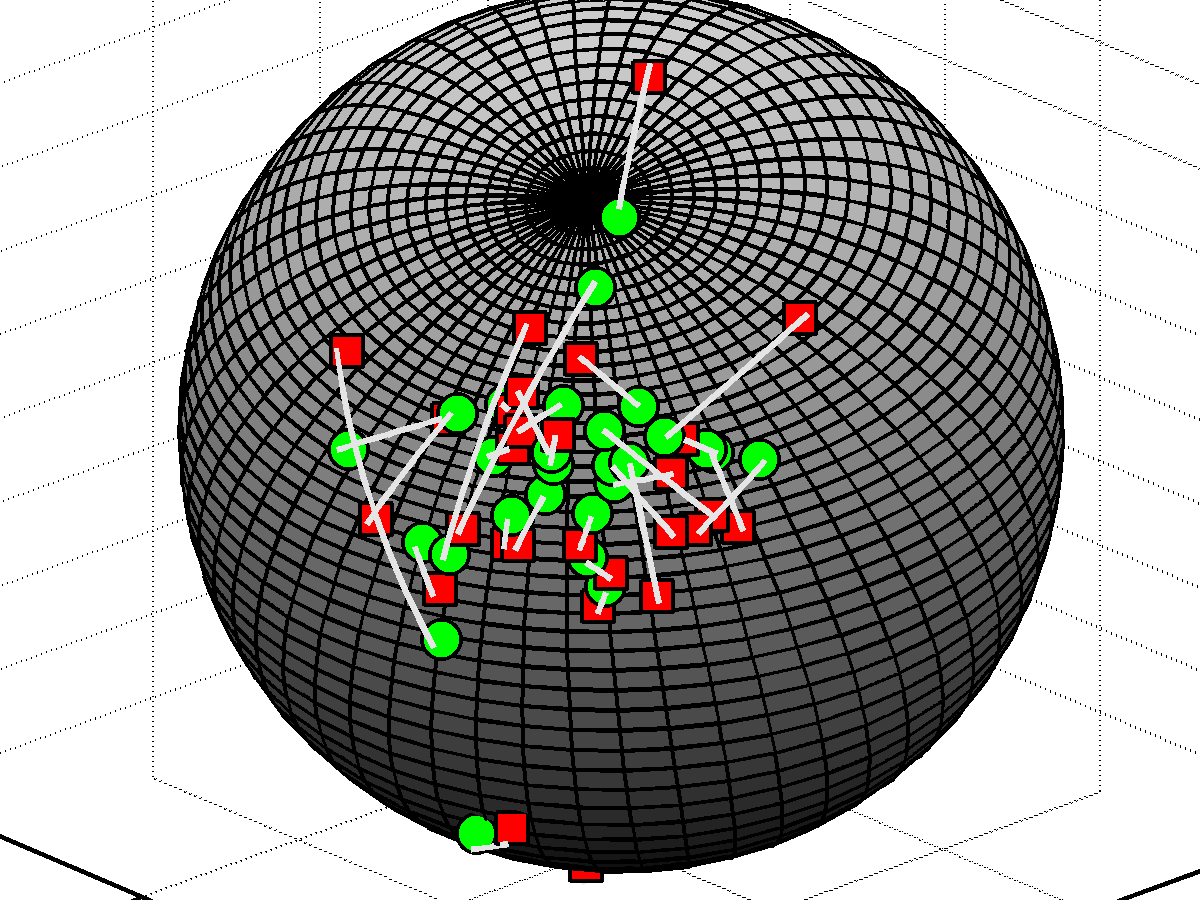}
&
\includegraphics[width=.4\textwidth]{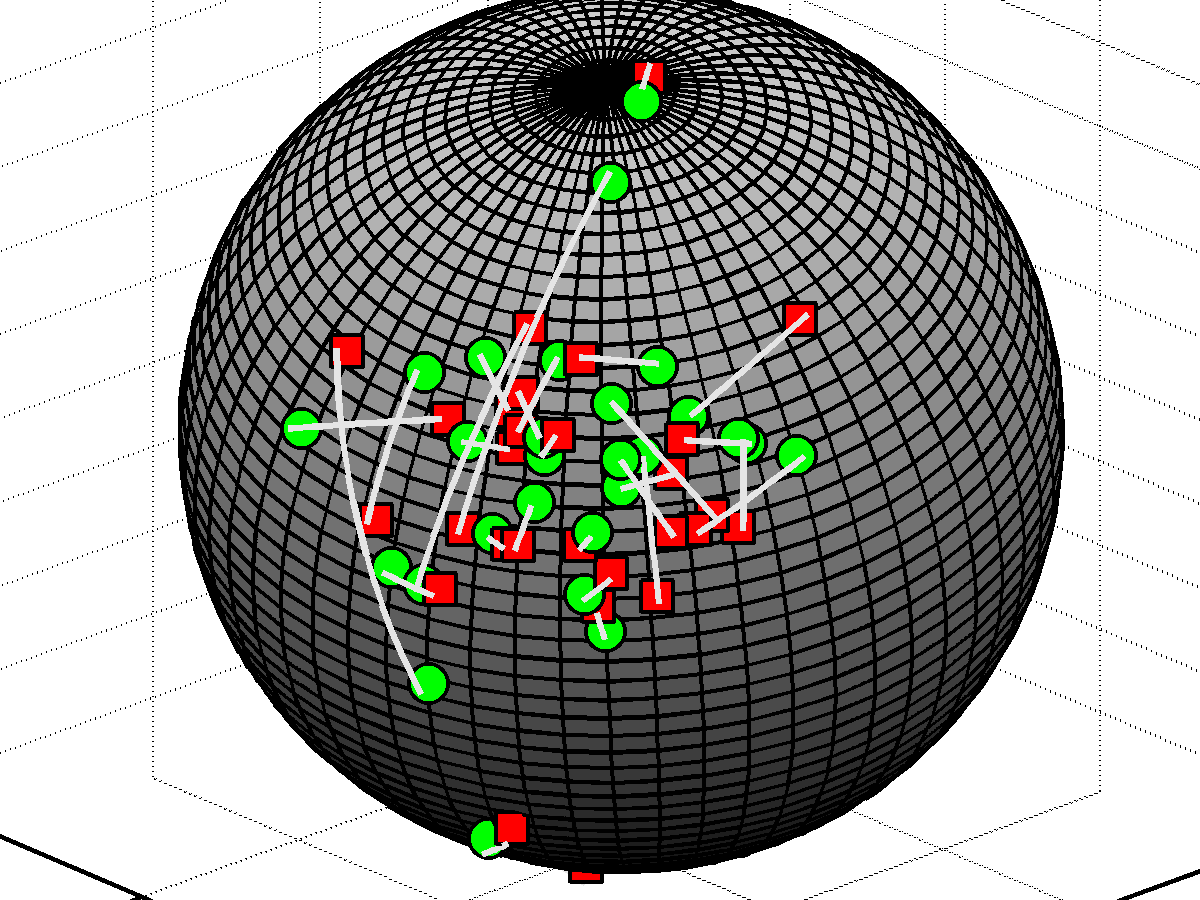}\\
\end{tabular}
\caption{Deformations of fitted models from vector-cardiogram training data with observed and predicted test data respectively plotted as red squares and green circles. Corresponding data points are connected with a light gray lines.  NLL refers to the nonparametric local linear regression model. PLT refers to the rigid rotation model. RR refers to the rigid rotation model.}
\label{fig:vectorcardiogram_pred}\end{figure}

The test and training data has been plotted on the sphere for each corresponding lead system from each child in Fig. \ref{figure:vector cardiogram}. The selected roughness parameter is $\lambda=10^{-2}$ and up to order $l=3$ is used for fitting the final model. A plot of the $MSE$ for the validation data used for tuning the model is in the supplementary. The third row of Table \ref{MSEtable} shows that our method has the smallest predictive error, while the previous nonparametric method proposed by \cite{Taylor:2013} has the largest prediction error. In Fig. \ref{fig:vectorcardiogram_pred}, the observed and predicted McFee directions are plotted for each model. 


\section{Discussion}


The asymptotic analysis is an important component of this method. This includes determining conditions necessary for consistency to occur. Unfortunately, the classical arguments for these types proofs usually involve many strong and subtle assumptions which are difficult to verify when analyzing over infinite dimensional nonlinear domains such as the group of diffeomorphisms on $\mathbb{S}^2$. To facilitate in bringing this problem forward we will outline the general argument typically used and point out the parts which we are uncertain about.

Several extensions to the proposed framework can result from alternative optimization strategies. This includes extensions into hyper-spherical domains. It is likely that one can extend the deformation basis utilized in this paper into hyper-spherical domains, 
however this may  be costly and inefficient. For higher dimensional spheres it may be better to construct custom orthonormal bases which are derived for specific applications in order to reduce the number of model coefficients. Additionally, one may wish to include a second order roughness penalty to the objective function. Other possible extensions include developing more general classes of deformations. For example, to better handle sliding plate boundaries one may wish to utilize a mapping that allows for discontinuity on sets of measure zero. 

\section*{Acknowledgements}
This research was supported in part by the NSF grants DMS 1208959 and IIS 1217515.

\section*{References}
\bibliography{bibfile}
\bibliographystyle{elsarticle-harv}

\appendix 
\section{Details of Numerically Computing the Roughness Measure}
\label{Computing_Roughness_Numerically}

We use polar coordinates to compute the roughness measure numerically. Given a point 
$(\theta,\phi)\in (0,\pi)\times (0,2\pi]$, 
we can obtain the cartesian coordinates as 
$\psi_{\theta,\phi}=
\left(
\sin(\theta)\cos(\phi),
\sin(\theta)\sin(\phi),
\cos(\theta)
\right)^T$.
Conversely, given a point $\p=(p_1,p_2,p_3)^T \in \s^2\subset\mathbb{R}^3$ in Cartesian coordinates, we can get its polar coordinates as 
$ \psi_\p^{-1}=(\theta,\phi)=(\cos^{-1}(p_3),\tan^{-1}(p_2/p_1))$.
For a warping $\gamma$ which is parametrized in Cartesian coordinates,  let $\z=(z_1,z_2,z_3)^T=\gamma(\psi_{\theta,\phi})$. The image of $\gamma$ at $(\theta,\phi)$ in
polar coordinates can thus be computed as
$(\tilde{\theta},\tilde{\phi})= (\cos^{-1}(z_3),\tan^{-1}(z_2/z_1))$. The variable $\phi$ can be visualized as a rotation about the $z$ axis and  $\theta$ can be thought of as the arc-length distance from the north pole. 

 To compute a Jacobian on $\gamma$, we take derivatives with respect to an orthonormal basis for each point on the sphere. An orthonormal basis for the tangent space at the point $(\theta, \phi)$ is computed as
$$
\left[
\frac{d\psi_{\theta,\phi}}{d\theta},
\frac{1}{\sin(\theta)}
\frac{d\psi_{\theta,\phi}}{d\phi}\right]
=\left[
\left(
\begin{array}{c}
\cos(\theta)\cos(\phi)\\
\cos(\theta)\sin(\phi)\\
-\sin(\theta)
\end{array}
\right),
\left(
\begin{array}{c}
-\sin(\phi)\\
\cos(\phi)\\
0
\end{array}
\right)
\right].
$$
Note that $\sin(\theta)=0$ at the north pole $\theta=0$ and south pole $\theta=\pi$. This implies that the change in distance on the sphere is zero at these two points when moving in the direction of $\phi$
, so
there are two points of discontinuity. Since we are integrating the roughness over the entire surface of the sphere, these two points will not theoretically change the measure of roughness. We exclude  
 small neighborhoods
around the north and south pole in our computation of the roughness.

Recall that the Jacobian operation is a linear mapping between the tangent spaces $T_{\theta,\phi}$ to $T_{\tilde\theta,\tilde\phi}$. Therefore, $J_{\theta,\phi}$ the Jacobian matrix with respect to the sphere at $(\theta,\phi)$ can be computed by applying a change of basis to the linear Jacobian matrix as
\begin{align*}
J_{\theta,\phi}&=
\left(
\begin{array}{cc}
1&0\\
0&\sin(\tilde{\theta})\\
\end{array}
\right)
\left(
\begin{array}{cc}
\frac{d\tilde{\theta}}{d\theta}&\frac{d\tilde{\theta}}{d\phi}\\
\frac{d\tilde{\phi}}{d\theta}&\frac{d\tilde{\phi}}{d\phi}\\
\end{array}
\right)
\left(
\begin{array}{cc}
1&0\\
0&\frac{1}{\sin(\theta)}\\
\end{array}
\right) =
\left(
\begin{array}{cc}
\frac{d\tilde{\theta}}{d\theta}&\frac{1}{\sin(\theta)}\frac{d\tilde{\theta}}{d\phi}\\
\sin(\tilde{\theta})\frac{d\tilde{\phi}}{d\theta}&\frac{\sin(\tilde{\theta})}{\sin(\theta)}\frac{d\tilde{\phi}}{d\phi}\\
\end{array}
\right)
\ .
\end{align*}
Looking at this mapping in reverse, the change of basis from the left hand side transforms the elements from the tangent space $T_{\tilde\theta,\tilde\phi}$ into coordinates with respect to $\e_1,\e_2$,
which is then applied to the linear Jacobian. Finally the matrix on the right transforms the basis back to elements of $T_{\theta,\phi}$.
A measure of the distance from isometry can be computed as $\|\logm(J_{\theta,\phi}^TJ_{\theta,\phi})\|$ where $\logm$ denotes the matrix log and $\|\cdot\|$ denotes the matrix Frobenius norm. 
The elements of the matrix $A_{\theta,\phi}=J_{\theta,\phi}^TJ_{\theta,\phi}$ can be computed as
\begin{align*}
A_{\theta,\phi}&=
\left(
\begin{array}{cc}
\frac{d\tilde{\theta}}{d\theta}
&
\sin(\tilde{\theta})\frac{d\tilde{\phi}}{d\theta}
\\
\frac{1}{\sin(\theta)}\frac{d\tilde{\theta}}{d\phi}
&
\frac{\sin(\tilde{\theta})}{\sin(\theta)}\frac{d\tilde{\phi}}{d\phi}
\\
\end{array}
\right)
\left(
\begin{array}{cc}
\frac{d\tilde{\theta}}{d\theta}&\frac{1}{\sin(\theta)}\frac{d\tilde{\theta}}{d\phi}\\
\sin(\tilde{\theta})\frac{d\tilde{\phi}}{d\theta}&\frac{\sin(\tilde{\theta})}{\sin(\theta)}\frac{d\tilde{\phi}}{d\phi}\\
\end{array}
\right)\\
&=
\left(
\begin{array}{cc}
a_{11}
&
a_{12}
\\
a_{21}
&
a_{22}
\\
\end{array}
\right)
=\left(
\begin{array}{cc}
\left(\frac{d\tilde{\theta}}{d\theta}\right)^2+ \sin^2(\tilde{\theta})\left(\frac{d\tilde{\theta}}{d\theta}\right)^2
&\hspace{1cm}
\frac{1}{\sin(\theta)}\frac{d\tilde{\theta}}{d\theta}\frac{d\tilde{\theta}}{d\phi}
+
\frac{\sin^2(\tilde{\theta})}{\sin(\theta)}\frac{d\tilde{\phi}}{d\theta}\frac{d\tilde{\phi}}{d\phi}
\\
\frac{1}{\sin(\theta)}\frac{d\tilde{\theta}}{d\phi}\frac{d\tilde{\theta}}{d\theta}
+
\frac{\sin^2(\tilde{\theta})}{\sin(\theta)}\frac{d\tilde{\phi}}{d\phi}\frac{d\tilde{\phi}}{d\theta}
&\hspace{1cm}
\frac{1}{\sin^2(\theta)}\left(\frac{d\tilde{\theta}}{d\phi}\right)^2
+
\frac{\sin^2(\tilde{\theta})}{\sin^2(\theta)}\left(\frac{d\tilde{\phi}}{d\phi}\right)^2
\\
\end{array}
\right)
\end{align*}

In this manner, we can measure the distance from isometry for the tangent space at each point by measuring the distance $J_{\theta,\phi}$ is from a rotation or reflection by
$
\| J_{\theta,\phi}^TJ_{\theta,\phi}-I_2\|=
\| A_{\theta,\phi}-I_2\|
$
where $I_2$ denotes the $2\times 2$ identity matrix and $\|\cdot \|$ denotes the Frobenius norm. This results in the
first-order roughness measure 
$Q(\gamma)= \int_{0}^{2\pi}\int_{0}^{\pi}\|A_{\theta,\phi}-I_2\|^2\sin(\theta)d\theta d\phi$.
Alternatively, the square of the Frobenius norm of the matrix log of $A_{\theta,\phi}$ can be computed using the eigenvalues as
\begin{align*}
\|\logm(A_{\theta,\phi})\|^2&= \log(\lambda_1)^2+\log(\lambda_2)^2 \mbox{, where }\\
\lambda_1&=\frac{a_{11}+a_{22}}
{2}+\frac{\sqrt{(a_{11}+a_{22})^2-4(a_{11}a_{22}-a_{12}a_{21})}}{2}\\
\lambda_2&=\frac{a_{11}+a_{22}}{2}-\frac{\sqrt{(a_{11}+a_{22})^2-4(a_{11}a_{22}-a_{12}a_{21})}}{2}\ .
\end{align*}
 This results in the following first-order roughness measure.
$R(\gamma)= \int_{0}^{2\pi}\int_{0}^{\pi}\|\logm( J_{\theta,\phi}^TJ_{\theta,\phi})\|^2\sin(\theta)d\theta d\phi$.

For a small $\delta>0$, let $\Theta$ denote a $M\times M$ matrix such that $\Theta_{i,j}=\pi(j-1+\delta)/(M-1-2\delta)$ for $i,j=1,\ldots, M$. Similarly, let $\Phi$ denote a $M\times M$ matrix such that $\Phi_{ij}=2\pi(i-1)/(M-1)$  for $i,j=1,\ldots, M$. In this manner, $\Theta$ and $\Phi$ represent a parametrization of the sphere where each point in the matrix can be mapped back to Cartesian coordinate using the mapping
$\psi(\Theta_{i,j},\Phi_{i,j})$. Let $(\tilde{\Theta}_{i,j},\tilde{\Phi}_{i,j})=\psi^{-1}(\psi(\Theta_{i,j},\Phi_{i,j}))$ for $i,j=1,\ldots, M$. One can discretely represent the warping function $\gamma\in \Gamma$ by these four matrices
$(\Theta, \Phi)$ and $(\tilde \Theta,\tilde\Phi)$. If the resolution $M$ is fine enough one can obtain numerically approximated derivatives for $(\tilde \Theta,\tilde\Phi)$. 
Since $\cos^{-1}$ and $\tan^{-1}$ are typically defined with range interval $[-\pi/2,\pi/2]$, care must be taken when computing the derivative numerically. To address this issue, let $m(a,b)= (a-b+ j\pi)$ with $j=\argmin_{i\in \mathbb{Z}}(|(a-b+ i\pi|)$.
A numerically approximated $2\times 2$ Jacobian matrix $J_{i,j}$ evaluated at $(\Theta_{i,j},\Phi_{i,j})$ can be computed as
\begin{align*}
J_{i,j}=
\left(
\begin{array}{cc}
\frac{m(\tilde \Theta_{i,j},\tilde \Theta_{i,j+1})}{m(\Theta_{i,j},\Theta_{i,j+1})}
&
\frac{m(\tilde \Theta_{i,j},\tilde \Theta_{i+1,j})}{m(\Phi_{i,j},\Phi_{i+1,j})}
\\
\frac{m(\tilde \Phi_{i,j},\tilde \Phi_{i,j+1})}{m(\Theta_{i,j},\Theta_{i,j+1})}
&
\frac{m(\tilde \Phi_{i,j},\tilde \Phi_{i+1,j})}{m(\Phi_{i,j},\Phi_{i+1,j})}
\\
\end{array}\right)
\end{align*}

\iftrue
A numerical estimate of $Q(\gamma)$is given by:
$$
Q(\gamma) \approx\sum_{i=1}^{M-1}\sum_{j=1}^{M-1}\trace((-I_2)^T(J_{i,j}^TJ_{i,j}-I_2))\sin(\Theta_{i,j}) 2 \pi^2/(M^2)\ .
$$
\fi
If we let $A_{i,j}=J_{i,j}^TJ_{i,j}$ and respectively denote the elements of the matrix as $a_{11,i,j}$,$a_{12,i,j}$,$a_{21,i,j}$, and $a_{22,i,j}$ then the eigenvalues can be computed as
\begin{align*}
\lambda_{1,i,j}=
\frac{a_{11,i,j}+a_{22,i,j}}
{2}+\frac{\sqrt{(a_{11,i,j}+a_{22,i,j})^2-4(a_{11,i,j}a_{22,i,j}-a_{12,i,j}a_{21,i,j})}}{2}\\
\lambda_{2,i,j}=
\frac{a_{11,i,j}+a_{22,i,j}}
{2}-\frac{\sqrt{(a_{11,i,j}+a_{22,i,j})^2-4(a_{11,i,j}a_{22,i,j}-a_{12,i,j}a_{21,i,j})}}{2}\ .
\end{align*}
A numerical estimate of $R(\gamma)$ is given by:
$$
R(\gamma) \approx\sum_{i=1}^{M-1}\sum_{j=1}^{M-1}
\left(\log\left(\lambda_{1,i,j}\right)^2+\log\left(\lambda_{2,i,j}\right)^2\right)
\sin(\Theta_{i,j}) 2 \pi^2/(M^2)\ .
$$

\section{Generating Arbitrary Diffeomorpism}
In the Sections 5.1 and 5.2, diffeomorphic maps were generated by composing several of a parametric diffeomorphisms.
The parametric families are: \\
\begin{itemize}
\item {\bf Rigid Rotation:} One can uniquely parametrize rigid rotations of the sphere using a special orthogonal matrix $R\in SO(n,\mathbb{R})$ which denote the set of $n\times n$ real valued matrices with $\det(R)=1$ and $R^TR=I_n$ where $I_n$ denotes the $n\times n$ identity matrix. A rotation diffeomorphism $\gamma$ which is parametrized via rotation matrix $R$ can be evaluated at each $z\in \s^2$ by the map $\gamma(z)=Rz$.
\item {\bf Projective Linear Transformation:} One can uniquely parametrize projective linear transformations using a special linear matrix $P\in SL(n,\mathbb{R})$ which denotes the set of $n\times n$ real valued matrices with $\det(P)=1$. A rotation diffeomorphism $\gamma$ which is parametrized via rotation matrix $R$ can be evaluated at each $z\in \s^2$ by the map $\gamma(z)=Pz/\|Pz\|$ where $\|\cdot\|$ denotes the Frobenius norm for matrices. This include the group of rigid rotations.
\item {\bf Conformal map:}
Consider $\s^2\subset \mathbb{R}^3$ and identify $\mathbb{R}^3$ with $\mathbb{C}\oplus \mathbb{R}$. 
So,  instead of writing an element of $S^2$ as $(x,y,z)$, we write it as $(z,t)$, where $z\in \mathbb{C}$ and $t\in \mathbb{R}$ and $|z|^2+t^2=1$.
For $\mathbb{S}^2$, the conformal maps are precisely the M\"obuis transformations studied by \cite{downs:2003}.
Suppose we are given the matrix $$M=\begin{pmatrix}a&b\cr c&d\cr\end{pmatrix}\in GL(2,\mathbb{C})\ .$$ Associated to this matrix is a conformal map $A:\s^2\to \s^2$, given by the following formula:
$$A(z,t)=
\left({2\overline{(cz+d(1-t))}(az+b(1-t))\over |az+b(1-t)|^2+|cz+d(1-t)|^2},{|az+b(1-t)|^2-|cz+d(1-t)|^2\over |az+b(1-t)|^2+|cz+d(1-t)|^2}
\right)
$$
This formula is well defined for every point of the sphere except the point where $z=0$ and $t=1$. At this point, $A$ is defined by 
$$A(0,1)=\left({2\overline{c}a\over |a|^2+|c|^2},{|a|^2-|c|^2\over |a|^2+|c|^2}\right)$$
Essentially this map is derived by conjugating a M\"obius tranformation by stereographic projection from $\mathbb{C}\to \s^2$. Note that every conformal map $\s^2\to \s^2$ is obtained by a matrix in this manner.\\

\item {\bf Twist map:}
Consider $\s^2\subset \mathbb{R}^3$ and identify $\mathbb{R}^3$ with $\mathbb{C}\oplus \mathbb{R}$. 
So,  instead of writing an element of $S^2$ as $(x,y,z)$, we write it as $(z,t)$, where $z\in \mathbb{C}$ and $t\in \mathbb{R}$ and $|z|^2+t^2=1$.
A very simple formula for a twist map is as follows. Start by fixing $r\in \mathbb{R}$. Then define a map $T:\s^2\to \s^2$ by
$$T(z,t)=(e^{irt}z,t)$$
Essentially this twist map takes each latitude circle to itself by a rotation, and this rotation varies from the south pole to the north pole.
\item {\bf Small Incremental Diffeomorphism Using Spherical Harmonic Basis:}
Let $B_1,B_2,\ldots, B_L$ denote the basis elements obtained from the spherical harmonics up to order $l$ as described in Section 4.2. One can parametrize a small incremental diffeomorphism which is close to the identity transformation via coefficient function $c \in \mathbb{R}^L$. If $\|c\|$ is sufficiently small then the map $z\in \s^2$ as $\gamma(z)=\exp_{z}(\sum_{i=1}^L c_iB_i(z))$ will represent a diffeomorphism which is close to the identity transformation. Since the set of diffeomorphism forms a group under composition, $\gamma_1 \circ \gamma_2$ is also a diffeomorphism for any two $\gamma_1,\gamma_2\in \Gamma$. In this manner one can get a larger diffeomorphism by iteratively applying these small incremental diffeomorphisms. Let $\gamma = \gamma_1 \circ \gamma_2\circ \ldots \circ \gamma_J$ denote $J$ compositions of small incremental diffeomorphisms.
\end{itemize}

The initialized deformation in Section 5.1 is generated from a composition of a rigid rotation parametrized by 
\begin{align*}
A_1= \left( \begin{array}{ccc}
    0.9564  &  0.2134 &  0.1994\\
   -0.2096  &  0.9770 &  -0.0403\\
   -0.2034  & -0.0032 &   0.9791\\
\end{array}
\right),\ 
\end{align*}
a projective linear transformation parametrized by
\begin{align*}
A_2= \left( \begin{array}{ccc}
    1.1874 &   0.4557 &   0.1407\\
    0.2148 &   1.0150 &   0.2162\\
    1.3649 & -0.2516  &  0.9063\\
\end{array}
\right),\ 
\end{align*}
a conformal transformation parametrized by
\begin{align*}
M= \left( \begin{array}{cc}
   0.8423 + 0.1561i & -0.0207 + 0.0537i\\
  -0.1746 + 0.0382i &  1.1054 - 0.0512i\\
\end{array}
\right),\ 
\end{align*}
and a twist map parametrized by $r=0.1088$.

The true deformation in Section 5.2 is generated from a composition of
a conformal transformation parametrized by
\begin{align*}
M= \left( \begin{array}{cc}
   0.8979 -0.23681i & -0.1256 + 0.2807i\\
  -0.3810 + 0.3379i &  0.6171 - 0.1121i\\
\end{array}
\right),\ 
\end{align*}
a twist map parametrized by $r=0.1877$, and a composition of 5 small incremental diffeomorphisms using up to order $l=2$ spherical harmonics. The 150 coefficients were generated randomly.

\section{Diffeomorphisms and Their Jacobean Eigenvalues}
Any diffeomorphism $\gamma: \mathbb{S}^2 \rightarrow\mathbb{S}^2$ by definition is differentiable at each point $p \in \mathbb{S}^2$. A map $\gamma$ is differentiable at $p \in \mathbb{S}^2$ if $J_{\gamma,p}$, the Jacobean of $\gamma$ at $p$, exists. If it is exists, then $J_{\gamma,p}$ is defined as the full rank linear map from $T_p(\mathbb{S}^2)$ to $T_{\gamma(p)}(\mathbb{S}^2)$ such that for each $v \in T_p(\mathbb{S}^2)$ the limit 
$$
J_{\gamma,p}(v)=\lim_{\epsilon\downarrow 0} \frac{\gamma(p+\epsilon v) -\gamma(p)}{\epsilon}\ 
$$
exists and yields a well defined linear map between two tangent spaces.
If the limit fails to exist for some $v \in T_p(\mathbb{S}^2)$, or maps to a lower dimensional subset of $T_{\gamma(p)}(\mathbb{S}^2)$, then we say that $J_p$ does not exist so that $\gamma$ is not differentiable at $p$. 

If $J_{\gamma,p}$ exists, then it can be represented as a full rank $2\times 2$ real valued matrix $B$ with respect to some choice of orthonormal bases respectively for $T_p(\mathbb{S}^2)$ and $T_{\gamma(p)}(\mathbb{S}^2)$. There is no standard basis to use so that the matrix representation will depend on the chosen basis. A change of basis will results in an orthogonal transformation $OBO^T$ for some $O\in \mathcal{O}(2)$ so that the Eigenvalues will not affected by the choice of basis. 

Let $\lambda_{1,p}$ and $\lambda_{2,p}$ denote the two Eigenvalues for $J_{\gamma,p}$. If $J_{\gamma,p}$ exists then both Eigenvalues must be finite and non-zero. Because $\gamma$ is a diffeomorphism, $J_{\gamma,p}$ is continuous with respect to $p$ so that the Eigenvalues $\lambda_{1,p}$ and $\lambda_{2,p}$ are also continuous with respect to $p$.

\section{Additional Plots and Results}
\subsection{Demonstration Using Simulated Data}
\begin{figure}[h!]\centering
\begin{tabular}{ccc}
Ours&NLL\\
\includegraphics[width=.4\textwidth]{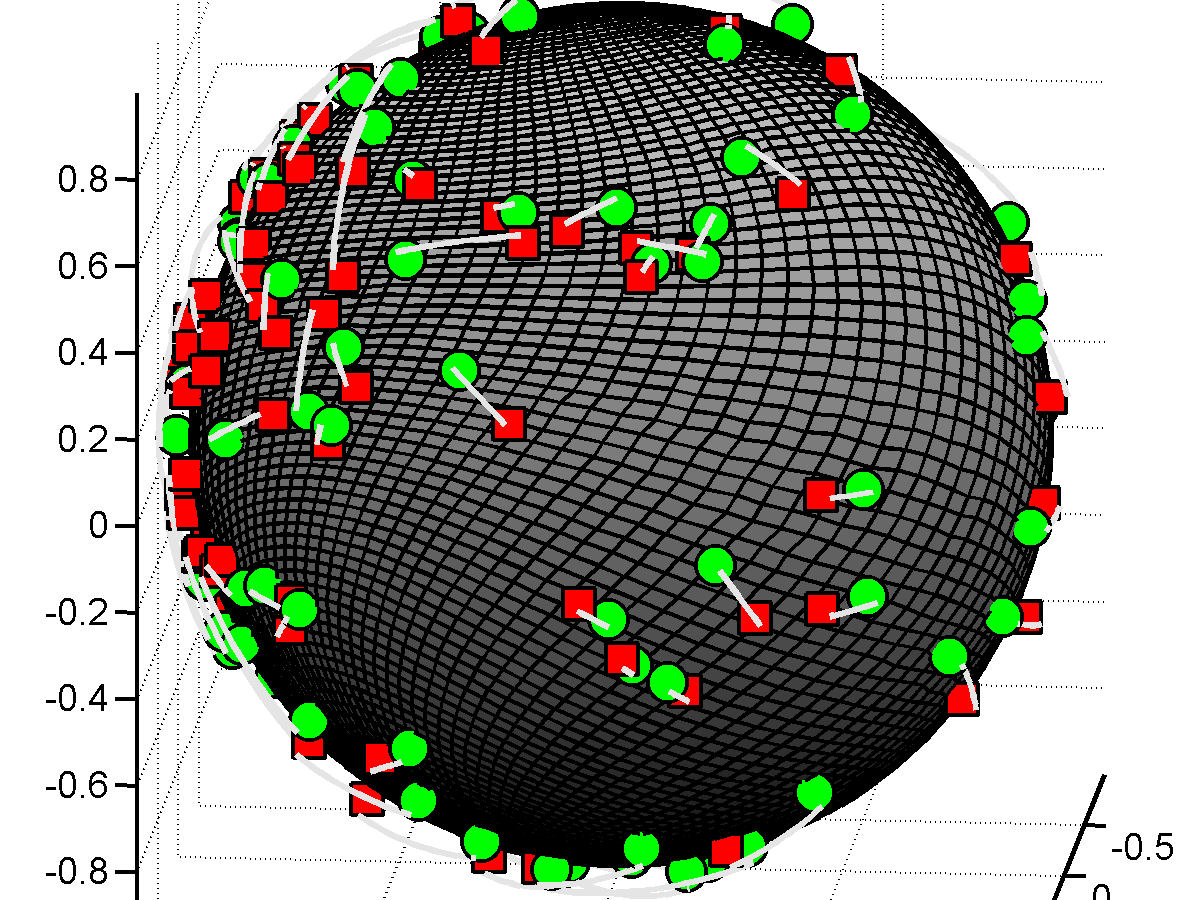}
&
\includegraphics[width=.4\textwidth]{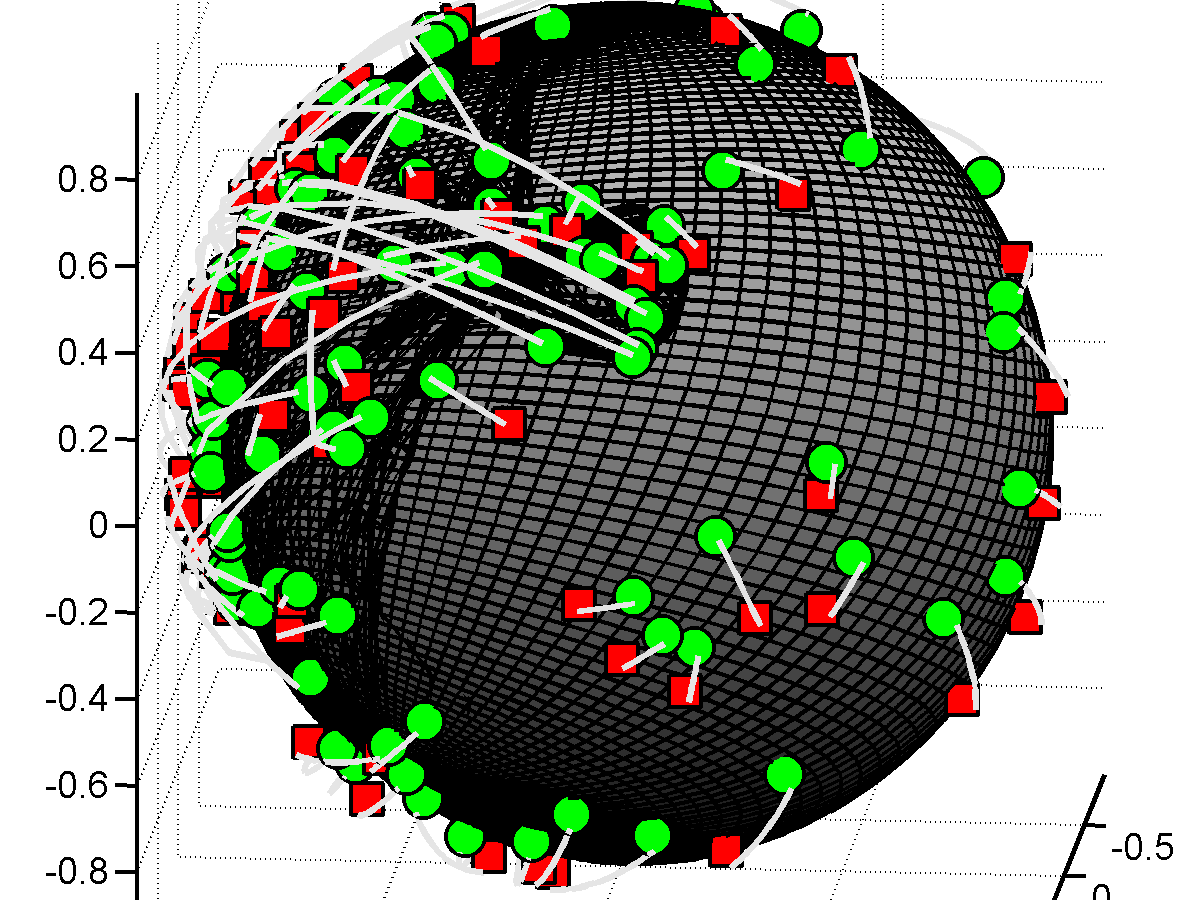}
\\
PLT&RR\\
\includegraphics[width=.4\textwidth]{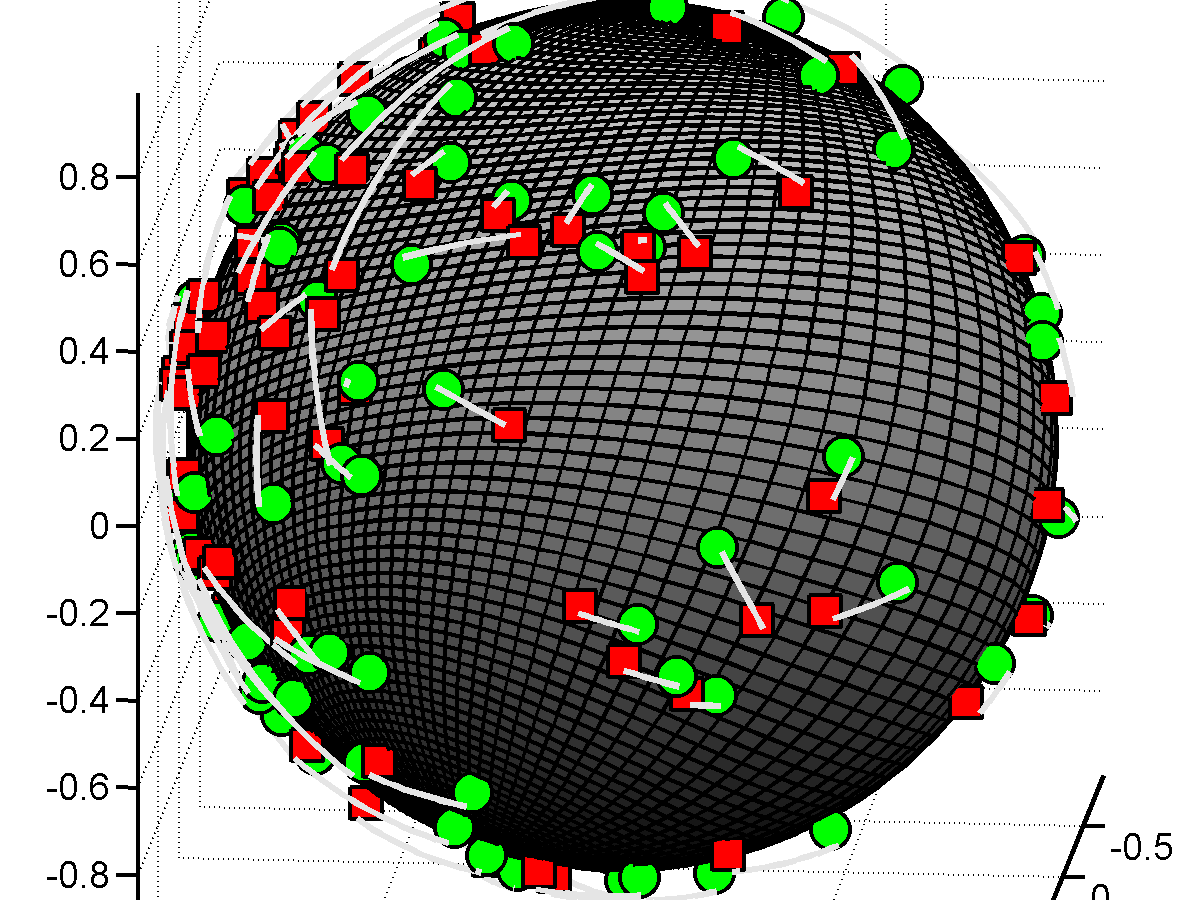}
&
\includegraphics[width=.4\textwidth]{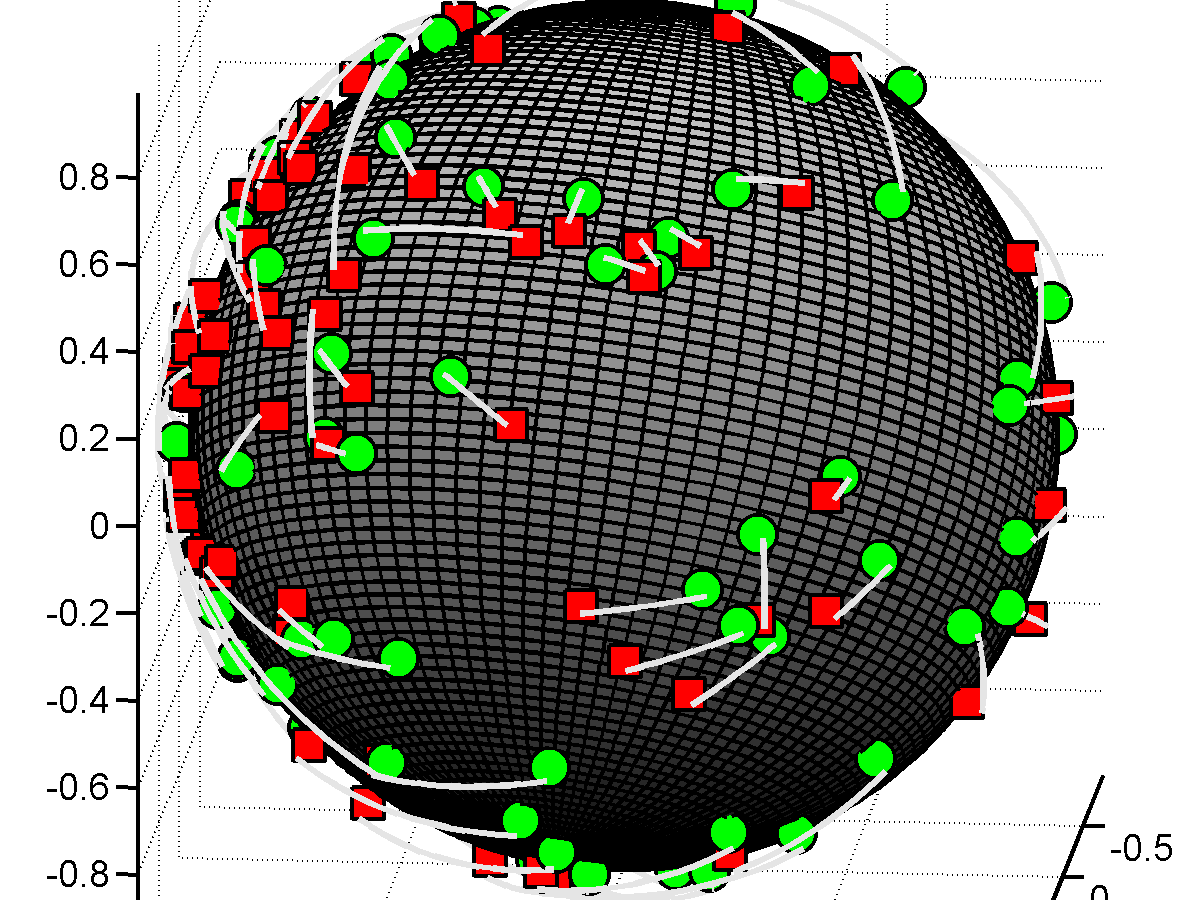}
\\
\end{tabular}

\label{fig:simdata_compair}
\caption{Deformations of fitted models from simulated training data with observed and predicted test data respectively plotted as red squares and green circles. Corresponding data points are connected with a light gray lines.
}
\end{figure}

Some additional plots from Section 5.2 which may be of interest are presented here. In Figure \ref{fig:simdata} the training and testing data are presented. In Figure \ref{fig:simdata_compair}, several other models fitted from the training data are compared. Notice the large overlapping regions that the nonparametric local linear model has in this case. This suggests that there is a problem with extrapolation here.
\begin{figure}[h!]\centering
\begin{minipage}[c]{.4\textwidth}\begin{center}
Training Data
\\
			\includegraphics[width=\textwidth]{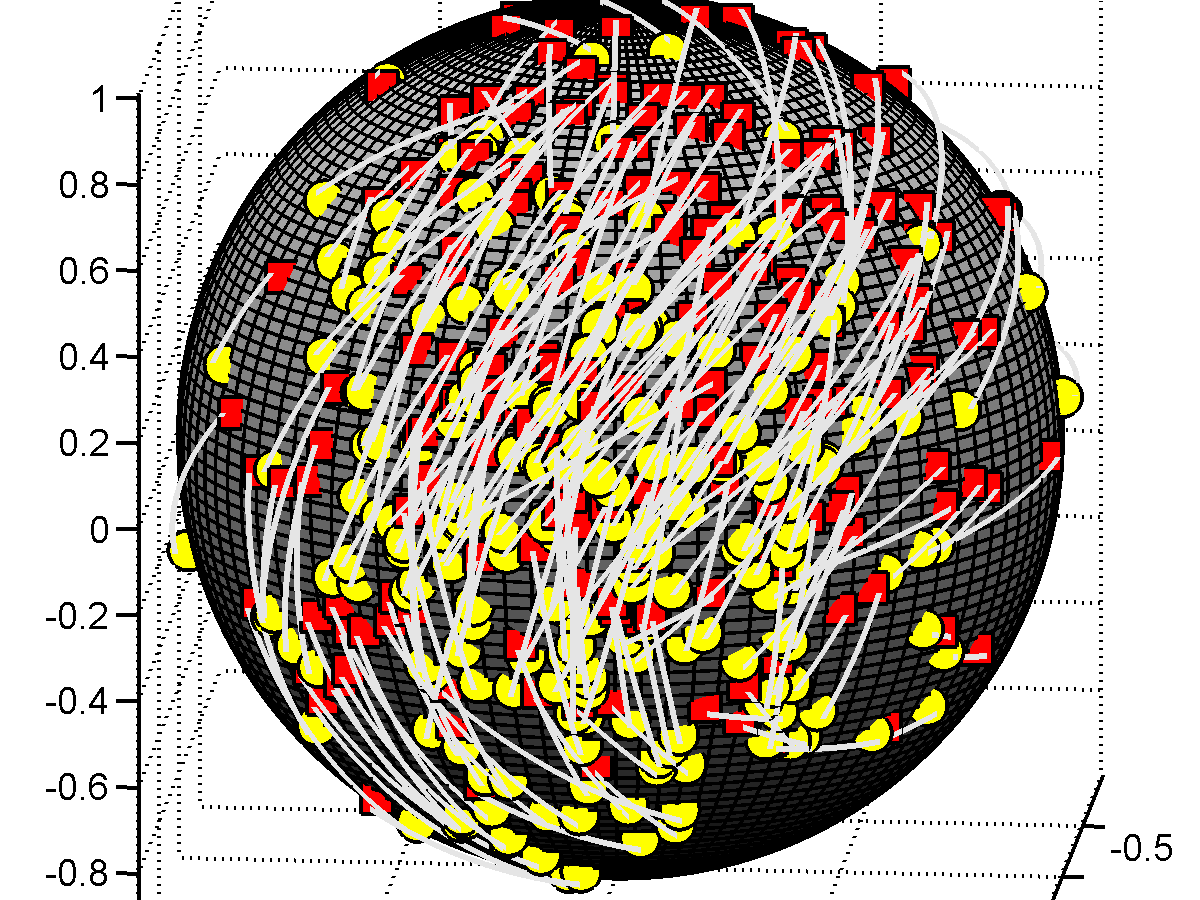}\\
			 ($n=200$)
			\end{center}
\end{minipage}
\begin{minipage}[c]{.4\textwidth}\begin{center}
Test Data\\
			\includegraphics[width=\textwidth]{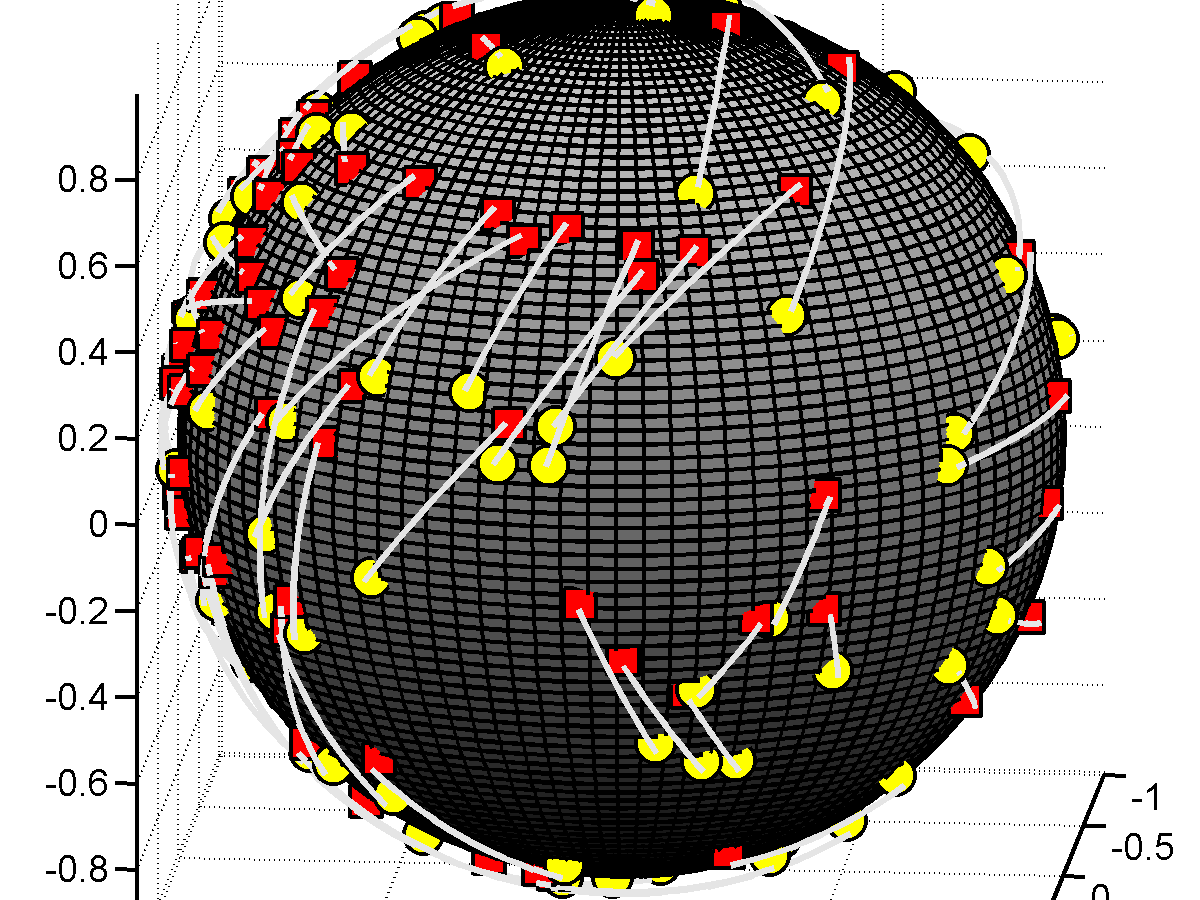}\\
			($n=100$)
			\end{center}
\end{minipage}
\caption{
The yellow points denote the $\x$'s and the red points denote the $\y$'s. Corresponding data points are connected with a light gray lines. 
}
\label{fig:simdata}
\end{figure}

\subsection{Weather Balloon Wind Directions Data}

\begin{figure}[h!]\centering
\begin{minipage}[c]{.4\textwidth}\begin{center}
$MSE$\\
			\includegraphics[width=\textwidth]{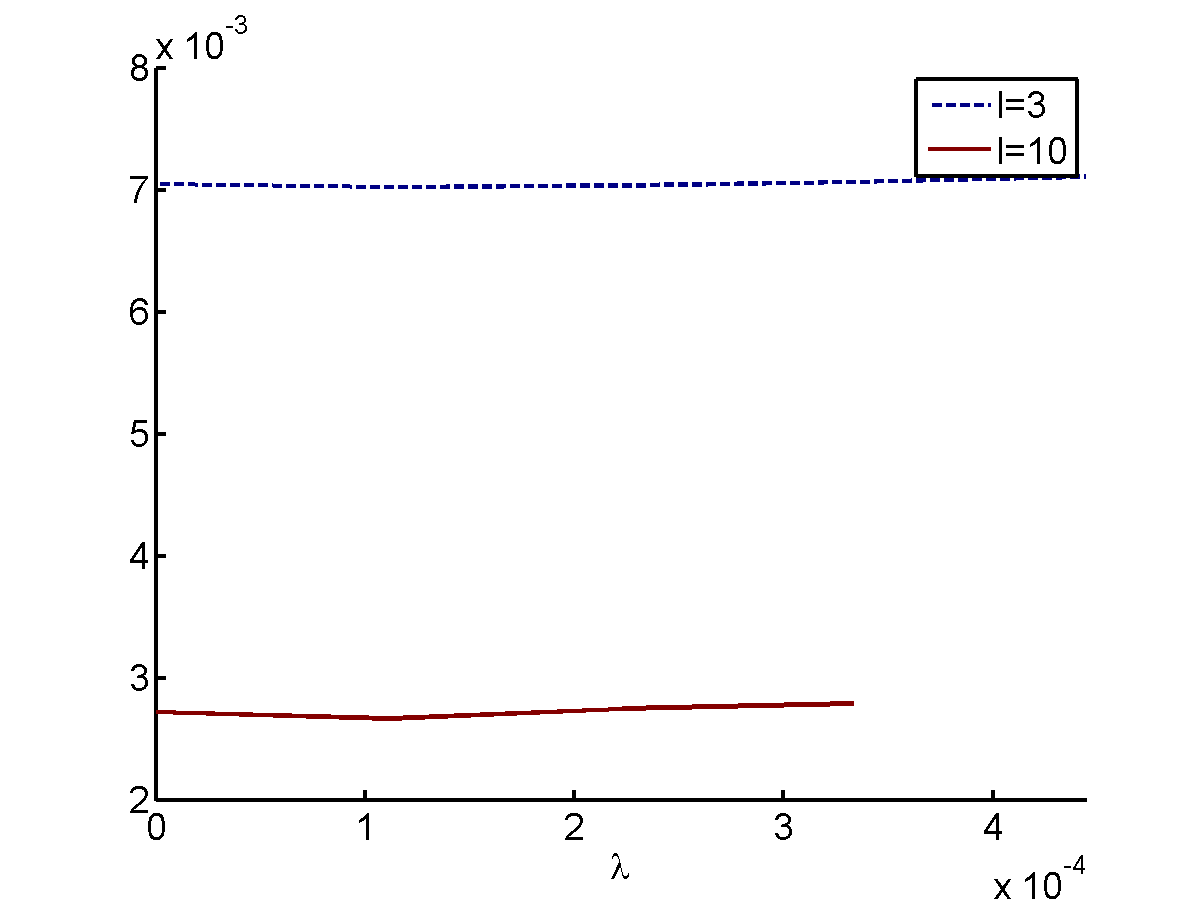}
			$\lambda$
			\end{center}
\end{minipage}

\caption{
One can see the mean squared error on the weather validation data for various values of $\lambda$. The dotted blue and solid red line respectively denote $l=3$ and $l=10$.
}
\label{fig:mseballoon}
\end{figure}

An additional plot from Section 5.3 which may be of interest are presented here. In Figure \ref{fig:mseballoon} one can see the validation error used for tuning the model. One can see that $l=10$ has a smaller validation error and is minimized around $\lambda=1.1200(10^{-4})$. Since the error is not greatly reduced with respect to $\lambda$ so we select $\lambda =10^{-10}$. There may be some higher frequency variability which might be better captured with a larger value of $l$ in this case.

\subsection{Vector-Cardiogram Data}

\begin{figure}[h!]\centering
\begin{minipage}[c]{.4\textwidth}\begin{center}
$MSE$\\
			\includegraphics[width=\textwidth]{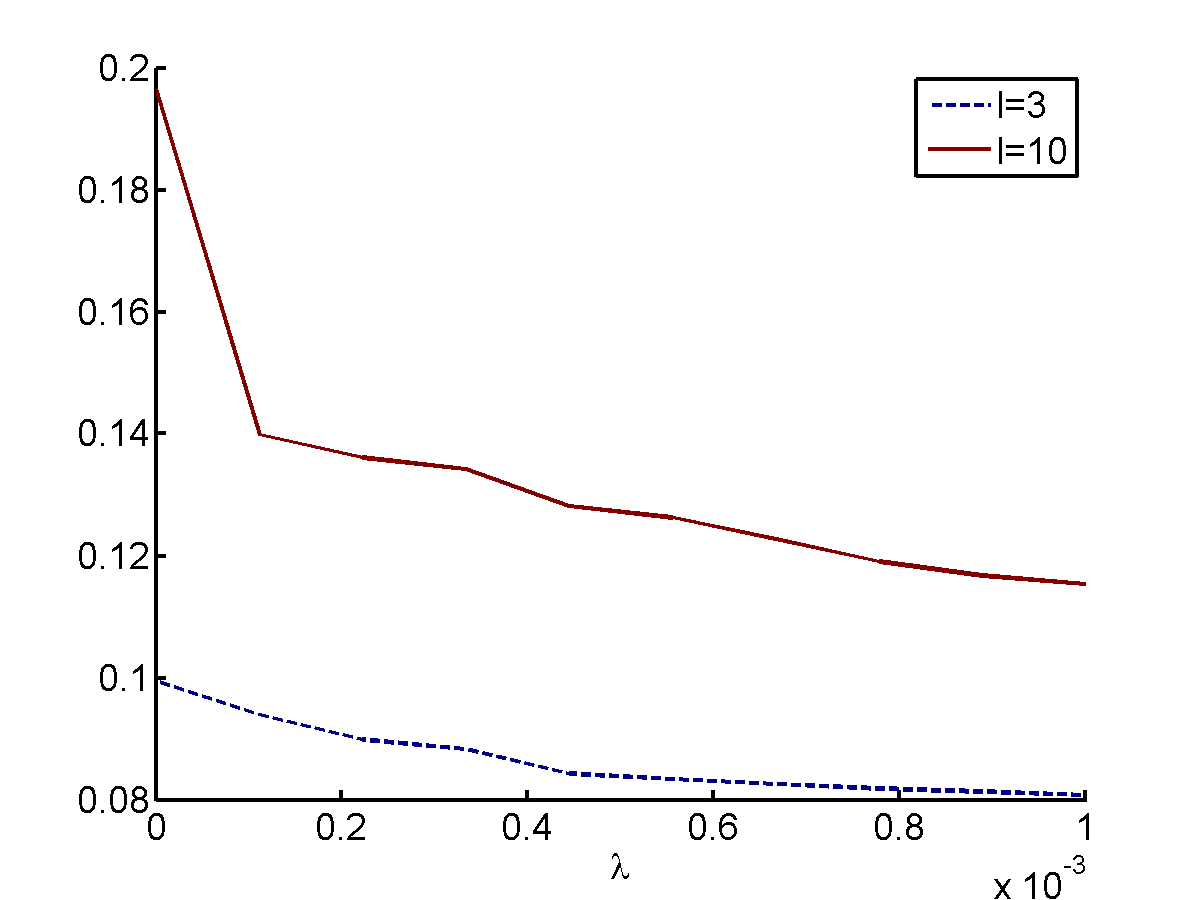}
			$\lambda$
			\end{center}
\end{minipage}

\caption{
One can see the mean squared error on the vector-cardiogram validation data for various values of $\lambda$. The dotted blue and solid red line respectively denote $l=3$ and $l=10$.
}
\label{fig:mseVector-Cardiogram }
\end{figure}

An additional plot from Section 5.4 which may be of interest are presented here. In Figure \ref{fig:mseVector-Cardiogram } one can see the validation error used for tuning the model. One can see that $l=3$ has a smaller validation error and is minimized with a heavy penalty. Since the roughness seems to be going down, we select $\lambda =10^{-2}$. The variability seems to be low frequency and may be better characterized using up to order $l=2$ basis elements.

\end{document}